\documentclass[fleqn,usenatbib]{mnras}

\usepackage{nccmath}    
\usepackage{graphics} 
\usepackage[dvipsnames]{xcolor} 

\usepackage[T1]{fontenc}

\DeclareRobustCommand{\VAN}[3]{#2}
\let\VANthebibliography\thebibliography
\def\thebibliography{\DeclareRobustCommand{\VAN}[3]{##3}\VANthebibliography}

\usepackage{graphicx}	
\usepackage{amsmath}	
\usepackage{amssymb}	
\usepackage{newtxtext,newtxmath}

\title[RGZ: GRG classification using Multi-Domain DL]{Radio Galaxy Zoo: Giant Radio Galaxy Classification using Multi-Domain Deep Learning}

\author[H.~Tang et al.]{
H.~Tang,$^{1,2}$\thanks{E-mail: hongming.tang@manchester.ac.uk}
A.~M.~M.~Scaife,$^{1,3}$
O.~I.~Wong$^{4,5,6}$,
S.~S.~Shabala$^{7}$
\\
$^{1}$Jodrell Bank Centre for Astrophysics, School of Physics and Astronomy, The University of Manchester, Manchester M13 9PL, England\\
$^{2}$Department of Astronomy, Tsinghua University, Beijing 100084, China\\
$^{3}$ The Alan Turing Institute, Euston Road, London NW1 2DB, UK \\
$^{4}$CSIRO Astronomy and Space Science, PO Box 1130, Bentley, WA 6102, Australia\\
$^{5}$ICRAR-M468, University of Western Australia, Crawley, WA 6009, Australia\\
$^{6}$ARC Centre of Excellence for All Sky Astrophysics in 3 Dimensions (ASTRO 3D), Australia\\
$^{7}$School of Natural Sciences, Private Bag 37, University of Tasmania, Hobart, TAS 7001, Australia\\
}

\date{Accepted XXX. Received YYY; in original form ZZZ}

\pubyear{2021}

\begin{document}
\label{firstpage}
\pagerange{\pageref{firstpage}--\pageref{lastpage}}
\maketitle

\begin{abstract}
In this work we explore the potential of multi-domain multi-branch convolutional neural networks (CNNs) for identifying comparatively rare giant radio galaxies from large volumes of survey data, such as those expected for new generation radio telescopes like the SKA and its precursors. The approach presented here allows models to learn jointly from multiple survey inputs, in this case NVSS and FIRST, as well as incorporating numerical redshift information. We find that the inclusion of multi-resolution survey data results in correction of 39\% of the misclassifications seen from equivalent single domain networks for the classification problem considered in this work. We also show that the inclusion of redshift information can moderately improve the classification of giant radio galaxies. 
\end{abstract}




\section{Introduction}
\label{sec:intro}

Radio galaxies are active galaxies with structures that typically consist of two radio lobes straddling a central host galaxy. These sources, together with radio loud quasars and some Seyfert galaxies, are often referred to as Double Radio sources associated with Active Galactic Nuclei \citep[DRAGNs;][]{1993LNP...421....1L}. The radio emission from DRAGNs is dominated by the synchrotron radiation process \citep{1955AZh....32..215S,1956ApJ...124..416B}, which allows one to calculate constraints on the strength of the local magnetic fields and the cosmic-ray energy spectrum of the emitting plasma \citep{1982phyn.book.....S}. They are believed to be powered by collimated, relativistic jets \citep{1974MNRAS.169..395B,1974MNRAS.166..513S} and magnetic fields of a certain geometry from their active nuclei \citep{2015aska.confE.109P}. Such active nuclei, or active forms of SuperMassive BlackHoles (SMBHs), have been found to reside in the most massive galaxies \citep{1982MNRAS.200..115S,1984ARA&A..22..471R,1984RvMP...56..255B,1998AJ....115.2285M,2013ARA&A..51..511K,2020A&A...642A.153D}. The radio lobes of DRAGNs then in turn can be used to probe the SMBHs in their host galaxy centres. To date, thanks to the availability of large scale radio sky surveys such as the Revised 3C catalogue \citep[3CR;][]{1959MmRAS..68...37E}, the Faint Images of the Radio Sky at Twenty-Centimeters \citep[FIRST;][]{1995ApJ...450..559B}, the NRAO VLA Sky Survey \citep[NVSS;][]{1998AJ....115.1693C}, \textcolor{black}{the Sydney University Molonglo Sky Survey \citep[SUMSS;][]{1999AJ....117.1578B,2003MNRAS.342.1117M}}, and the Westerbork Northern Sky Survey at 325\,MHz \citep[WENSS;][]{1997A&AS..124..259R}, tens of thousands of radio galaxies have been identified.

\begin{figure}
\setlength{\unitlength}{1cm}
\includegraphics[width=0.45\textwidth]{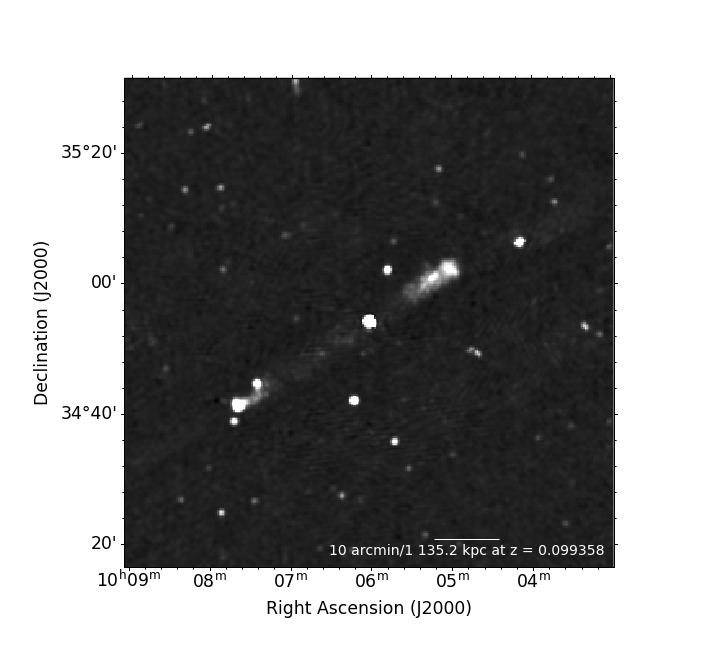}
\caption{The NVSS radio map of 3C\,236, a GRG example. The map is shown in logarithmic scale, with the scale bar on the diagram represents 10 arcmin in angular size. This is consistent to the linear scale of 1,135.2\,kpc at z = 0.099358 \citep{2006PASP..118.1711W}.}
\label{fig:3C_236}
\end{figure}

The large statistical samples from these sky surveys have motivated research into the maximum size to which a radio source might evolve, resulting in the discovery of the first two Giant Radio Galaxies \citep[GRGs;][]{1974Natur.250..625W}. GRGs are defined as those radio galaxies that have a projected linear size greater than 700\,kpc \citep{dabhadelotss} under a $\rm \Lambda$CDM cosmology with $\Omega_{\rm m}$ = 0.31 and a Hubble constant of $\rm H_{\rm 0} = 67.8~km~s^{-1}~Mpc^{-1}$ \citep{2016A&A...594A..13P}. \textcolor{black}{For example,  Figure~\ref{fig:3C_236} shows a log-scale radio map of 3C\,236 from NVSS at 1.4\,GHz, one of the first two GRGs discovered. The source has a host spectroscopic redshift of 0.099358, measured from the Sloan Digital Sky Survey \citep[SDSS;][]{2017ApJS..233...25A},  and therefore an  angular size distance of 
1.892\,kpc/arcsec \citep{2006PASP..118.1711W}. Since the end-to-end angular extent of the source as estimated from the NVSS map is 2505\,arcsec, the estimated physical linear size of 3C~236 is 4.7\,Mpc, significantly larger than the 700\,kpc limit in the GRG definition.}

The primary motivation for finding such giant radio sources is to investigate the possible modes of energy replenishment that allow for the existence of such a population \citep{1973MNRAS.164..243L}, as energy losses over the physical scales associated with these gigantic radio components are unavoidable. Furthermore, such objects can also be used as probes of the local intergalactic medium (IGM), as their structures are likely to be  influenced by their environment, and  it has been shown in particular that GRGs can be used to investigate the missing baryon problem in inter-galactic filaments, as probes of the Warm Hot Intergalactic Medium \citep[WHIM;][]{2015aska.confE.109P}. 

\cite{1974Natur.250..625W} have also pointed out that the large angular extent of these radio sources allows detailed imaging of such structures, and can therefore assist in studies of the physical processes occurring within the galaxies themselves. For instance, \citet{2019MNRAS.482..240K} investigated well-resolved radio maps of 33 3CR radio sources. They found that 24 objects in their sample showed strong jet precession evidence, which is consistent with the hypothesis of black hole binary mergers \citep[e.g.][]{2013ApJ...768...11B,2018A&A...617A..58C}. This idea has also motivated an interest in the discovery of GRGs with unusual morphologies. For instance, by discovering several giant Double-Double Radio Galaxies, \citep[DDRG;][]{2000MNRAS.315..371S,2006MNRAS.366.1391S,2014ApJ...788..174B,2016MNRAS.460.2376B}, it was found that jet-interruption might have taken place within radio sources with such unique radio morphology. In another example, \cite{2011AstBu..66..416S,2014AstL...40..606S} reviewed a list of GRG candidates, and found 8 radio sources with signatures of galaxy interaction. These objects have X-shape radio morphologies, and are believed to be in the final stage of mergers. To date, there are over 800 GRGs identified in the literature  \citep{kuzmicz2018,2020A&A...642A.153D,2020A&A...642A.153D,2020MNRAS.499...68T,2021MNRAS.501.3833D}, and the work of hunting for new GRGs is still ongoing.

In this paper we introduce a novel automated method for classifying radio galaxies as giants or non-giants based on a convolutional neural network approach. The structure of the paper is as follows: in \S~\ref{sec:classification} we review existing approaches to the classification of giant radio galaxies; in \S~\ref{sec:algorithm_theory} we explain the theoretical background to the algorithms used in this work; in \S~\ref{sec:data_sample} we describe how we selected and built two machine learning data sets with different object class ratios and sample constitutions; in \S~\ref{sec:network_architecture} and \S~\ref{sec:discussion} we compare the model performances resulting from different training strategies and data samples, including an interpretation of different model behaviour and the connection to data selection; we finally draw our conclusions in \S~\ref{sec:conclusions}.

In this work we assume a $\rm \Lambda$CDM cosmology with $\Omega_{\rm m}$ = 0.31 and a Hubble constant of $\rm H_{\rm 0} = 67.8~km~s^{-1}~Mpc^{-1}$ \citep{2016A&A...594A..13P}. The model training, validation and testing are performed using the Google Colaboratory \citep{Bisong2019}, which is equipped with a NVIDIA Tesla T4 (14.7\,GB memory).

\section{GRG Classification Approaches}
\label{sec:classification}

Regardless of exact methodology, classifying a radio galaxy as a GRG requires astronomers to (i) identify radio components belonging to the same DRAGN; (ii) measure the source's Largest Angular Size (LAS) from radio maps; (iii) find the corresponding host galaxy of the DRAGN; (iv) measure the host galaxy redshift, and (v) derive the source's projected physical linear size based on the source LAS and  host galaxy redshift. In the following section, we recap previous GRG classification methodologies, and highlight the advantages and disadvantages that have motivated the algorithm development presented in this work.

\subsection{Visual inspection}

The majority of historic GRG studies use `by eye' classification, also known as visual inspection. In these studies, new GRGs were confirmed by following-up sample candidates pointed out by previous studies, or by searching large-scale radio survey catalogues by eye \citep{1974Natur.250..625W,1976Natur.262..179B,1977MNRAS.181..465W,1983MNRAS.204..151L,1986A&A...169...63K,1989A&A...226L..13D,1989PASAu...8...81J,1989MNRAS.236..737E,1993MNRAS.264..721L,1995MNRAS.277..995L,1996MNRAS.281.1081C,1996ApJS..107...19M,1996MNRAS.279..257S,1999MNRAS.309..100I,2000A&AS..146..293S,2001A&A...378..826L,2001A&A...371..445M,2002MNRAS.329..227S,2004A&A...424..455L,2005AJ....130..896S,2006MNRAS.366.1391S,2008ApJ...679..149M,2007AJ....133.1331H,2008ApJ...679..149M,2011MNRAS.415.1013K,2011MNRAS.417L..36H,2011AstBu..66..416S,2014A&A...565A...2M,2014AstL...40..606S,2014ApJ...788..174B,2015AstBu..70...45A,2015MNRAS.453.2438T,2016AstBu..71..384A,2017MNRAS.469.2886D,2017A&A...601A..25C,2017AJ....154..253K,2018MNRAS.480..707P,2018MNRAS.473.4926S,2020yCat..22470053K,dabhadelotss,2020A&A...642A.153D,2020MNRAS.499...68T,2020arXiv201205759D}. 

Before large scale radio surveys such as NVSS or FIRST were available, most studies in this field would examine the validity of a GRG candidate by making a deep observation of a particular source or set of sources in specific radio and optical wavebands and would use the optical spectrum of the galaxy host in order to measure source redshift \citep[e.g.,][]{2014ApJ...788..174B,2015AstBu..70...45A,2015MNRAS.453.2438T}. When large-scale radio surveys (e.g. NVSS, SUMSS, FIRST), \textcolor{black}{and optical surveys such as SDSS became available}, later studies tended to select source candidates from a particular survey and to perform early cross validation using other survey image data if available. With the availability of photometric redshifts for large numbers of objects from surveys such as SDSS, a number of more recent discoveries have also used photometric redshift when estimating object distances \citep[e.g.,][]{dabhadelotss,2020A&A...642A.153D,2020MNRAS.499...68T}. Such approaches have enabled researchers to measure source host redshift and object LAS of these GRGs with excellent reliability, and investigated their spectral properties \citep[e.g.,][]{dabhadelotss}.

In addition to experts, citizen scientists have also recently joined the hunt for GRGs. Radio Galaxy Zoo \citep[RGZ;][]{2015MNRAS.453.2326B}, an online citizen science project, aims to cross match radio sources at 1.4\,GHz with their infrared host galaxies. Although the project was initially launched to create a large scale radio galaxy catalogue with the help of citizen scientists, its online forum \emph{RadioTalk} allowed citizen scientists to collaborate with project team scientists and find radio galaxies of special types. Four out of the six radio galaxies confirmed as GRGs by the RGZ team were pointed out on the forum as GRG candidates by several project citizen scientists in advance of their confirmation \citep{2016MNRAS.460.2376B,2020MNRAS.499...68T}.

\subsection{Automated Searches}
\label{sec:automated}

Each of the GRG identification processes described in the previous section relies largely on visual inspection and manual analysis. These methods are widely accepted as they allow for cross validation with diverse complementary radio/optical survey images, deep source imaging and/or spectral confirmation. The consensus level of such approaches is therefore extremely high. However, while such approaches work well when dealing with source catalogues such as NVSS, SUMSS and FIRST, with modest sample sizes, such methods become impractical when faced with millions rather than thousands of candidate galaxies. For example, the Evolutionary Map of the Universe \citep[EMU;][]{2011PASA...28..215N} survey, one of the the Australia SKA Pathfinder \citep[ASKAP;][]{2008ExA....22..151J} early science projects, is expected to provide a catalog containing approximately 70 million sources \citep{2011PASA...28..215N,2021PASA...38...46N}, of which $\sim 7$ million will require visual inspection. Reliable automated GRG search methods therefore become necessary in the era of astronomical big data.

One recent GRG search method \citep{2016ApJS..224...18P} attempted to challenge the traditional visual inspection methods by using a decision tree based machine learning approach. The study focused on sources with only two radio components and aimed to identify GRGs among them, regardless of component morphology (lobes, jets, core). The training set for this work consisted of 51\,195 source pairs from the NVSS catalogue, of which 48 had previously been confirmed to be GRGs by \citet{2001A&A...378..826L}. When selecting sample features, the study mostly used component major axis, minor axis and peak flux as input features. The best classifier in this study achieved a training accuracy of 97.8$\pm$1.5\%. \citet{2016ApJS..224...18P} then used the best classifier to find GRGs from 870\,000 candidate pairs in the full NVSS catalogue, extracting those objects with high GRG probabilities. Such a semi-automated procedure produced a list of 1\,616 GRG candidates with LAS $\ge$ 4$'$.

Although this pioneering study predicted a large number of GRG candidates, it did not consider source host galaxy redshift as this was unavailable in the NVSS catalogue. Consequently, only 16 of the selected candidates were included in the later updated catalogue of GRGs by \citet{kuzmicz2018}. This catalogue assumes $\rm H_{0}$ = 71\,$\rm km\,s^{-1}\,Mpc^{-1}$, $\Omega_{M}$=0.27, $\Omega_{\rm vac}$=0.73, and lists 349 confirmed GRGs as of the end of 2018. The catalogue uses host galaxy redshifts either from the literature or from SDSS. 

Limited by the lack of host galaxy redshift information, validation of the sample from \citet{2016ApJS..224...18P} was not comprehensively addressed until \citet{2020A&A...642A.153D} performed a follow up study. Thanks to the availability of source coordinates in the \citet{2016ApJS..224...18P} candidate catalogue, the team was able to track positions for each candidate, manually visualize their NVSS, FIRST, The GMRT 150 MHz All-sky Radio Survey \citep[TGSS;][]{2017A&A...598A..78I}, and the Karl G. Jansky Very Large Array Sky Survey  \citep[VLASS;][]{2020PASP..132c5001L} images (if available), and also check their host galaxy redshift from publicly available optical surveys and databases \citep{2020A&A...642A.153D}. Source LAS were measured from NVSS images for uniformity, where only  emission above a 3$\sigma$ level was considered. The team found that there were 165 known and 151 newly discovered GRGs among the candidates. In other words, around 20.8\% of the candidates in the list were finally confirmed to be GRGs. 

Although the algorithm of \citet{2016ApJS..224...18P}  did not predict a fully reliable set of GRGs from the test sample, it had successfully produced a good candidate pool. The availability of traceable source coordinates allowed follow up work to perform traditional visual inspection and cross validation. Such availability further allowed them to check sample host galaxy redshifts, as source LAS and redshift are the two key factors to determine whether a source is a GRG. As a result, the \citet{2016ApJS..224...18P} candidate list has contributed more than 35\% of all sources to the total confirmed GRG popularization.

In spite of its success, the algorithm presented in  \citet{2016ApJS..224...18P} also raises questions about selection biases: for example, in this instance only sources with two radio components and large LAS ($\ge$4$'$) are considered. In other words, the \citet{2016ApJS..224...18P} algorithm considered only those radio sources with a host galaxy redshift larger than $z=0.17$. Such selection biases would have excluded at least 108 GRGs of smaller LAS in the \citet{kuzmicz2018} catalogue. Also, it is problematic that GRGs with more complicated morphologies could not be recognized by the \citet{2016ApJS..224...18P} algorithm as these are considered important for particular types of investigation, as described in Section~\ref{sec:intro}. In light of these considerations, in this work we present a GRG classifier capable of identifying GRGs of smaller LAS and with diverse radio morphologies using an approach based on Convolutional Neural Networks \citep[CNN;][]{NIPS2012_4824}.  Considering the traditional approaches to GRG candidate validation using multi-frequency radio survey data, we also explore the possibility of using multi-survey image data and host galaxy redshifts as algorithm inputs.

\section{Convolutional Neural Networks}
\label{sec:algorithm_theory}

In recent years, CNNs have become widely used for astronomical image pattern recognition related problems, such as galaxy cluster and filament detection \citep{2018MNRAS.480.3749G}, supernovae classification/detection \citep[e.g.,][]{2017arXiv171111526K,2020PhRvD.102d3022C}, compact and extended radio galaxy distinction \citep{2018MNRAS.476..246L}, and radio galaxy localization \citep{2019MNRAS.482.1211W}. CNNs are popular as they can decompose 1-D or 2-D inputs into partly overlapped patches, where each cortex neuron only captures features from a specified patch \citep{MATSUGU2003555,2019arXiv190503554K}. Also, CNNs are weight sharing, allowing them to be translation equivariant and therefore robust to offsets in radio images.

In the context of radio galaxy morphology classification, most state-of-the-art CNN applications have concentrated on single image inputs. Specifically, both training and test samples typically come from only one radio sky survey \citep{2017ApJS..230...20A,2018MNRAS.476..246L,2018MNRAS.480.2085A,2019MNRAS.488.3358T,2021MNRAS.501.4579B,e2cnn}, or have radio source contours overlaid on infrared sky survey maps \citep{2018MNRAS.478.5547A,2019MNRAS.482.1211W}. \textcolor{black}{Notably, \citet{2018MNRAS.478.5547A} trained their single image input CNN along with 10 derived features to cross-identify radio galaxies and their corresponding host galaxies. Although they themselves did not highlight it in their text, this algorithm could be regarded as the first multi-domain CNN in the field of radio galaxy classification.}

\textcolor{black}{More broadly in astronomy, AstroNet \citep[i.e.][]{2018AJ....155...94S} is the earliest and the most well known multi-branch CNN \citep{2017arXiv171005477L} application. AstroNet is multi-branch neural network which is used to find exoplanet candidates from light curves observed by NASA's {\tt Kepler Space Telescope} \citep{2010ApJ...713L..79K}. The authors imported both `global view' (fixed-length complete light curve) and `local view' (fixed-length window over the detected transit) data of the observed exoplanet light curve into their algorithm, and successfully identified two new exoplanet candidates. Following AstroNet, several other deep-learning applications \textcolor{black}{also used the mutli-branch strategy} have been made to detect exoplanets \citep[i.e.][]{2018ApJ...869L...7A,2020A&A...633A..53O} or Fast Radio Bursts \citep[FRB; i.e.][]{2018AJ....156..256C}. }

Considering the conventional procedure of GRG identification, the proposed algorithm needs to be able to learn from multiple radio survey images, which potentially could be achieved by the aforementioned multi-branch or multi-domain networks. These algorithms could ideally concatenate the image features learned from the convolutional layers with parameterised features such as host galaxy redshift and provide a combined input to a subsequent model layer (e.g. a fully-connected layer). Such CNN based algorithms that are able to concatenate features learned from different inputs are referred to as multi-domain multi-branch CNNs . 

\subsection{Multi-branch CNNs}
\label{sec:MultiBranNN}

\textcolor{black}{Concepts similar in nature to that of multi-branch neural networks can be found from 2016, when \citet{2016arXiv160607792C} announced an algorithm for recommender systems based on feed-forward neural networks. They referred to their approach as `Wide and Deep Learning', where `wide' refers to wide linear models and  `deep' represents deep neural networks. Similar approaches using CNNs as a backend soon appeared, including \citet{2017arXiv170601788A}, which attempts to merge two independent top-down CNNs.} That work initially developed a network architecture that they referred to as a multi-domain CNN to localize JPEG double compression. In their model, a spatial domain CNN (2-D input) and a frequency domain CNN (1-D input) had their outputs passed to two fully-connected layers, the outputs from which were then concatenated into a larger fully connected layer. The outputs from this layer  then served as an input to subsequent layers. Such an architecture allowed the model to learn all features jointly through end-to-end back-propagation and it was shown that such approach could provide superior model performance for specific problems compared to a traditional spatial domain CNN \citep{2017arXiv170601788A}.

Inspired by the \citet{2017arXiv170601788A} architecture, several multi-branch CNN algorithms have subsequently been developed and are being used in computer vision and medical research \citep{2017arXiv171005477L,2018arXiv181102942A,2018IJSTA..11.4299S,2019arXiv190508413C}. Thanks to the availability of feature concatenation, these studies were able to do bulk model training on different image inputs, different channels of the same image data cube, or have each branch trained on the same image data using different kernel sizes.  Multi-branch networks that use multiple inputs from different sources, such as 2-d image data and 1-d spectral data, are also referred to as \emph{multi-domain} CNNs as they perform joint learning on data from different domains.

In the context of GRG classification, it is essential for an algorithm to be able to estimate source LAS from 2-D image inputs and to also use the source host redshift as a scalar feature in order to link the LAS to the physical size of the object.  Thanks to the concatenation feature of multi-branch CNNs, such multi-domain inputs can be joined together in a bulk training process either from the top of their corresponding branches or simply from the concatenation layer. This makes multi-branch CNNs an ideal candidate algorithm for developing automated GRG object classifiers. In the following section we describe the multi-domain data set constructed for use in this work and in Section~\ref{sec:network_architecture} we describe the multi-branch model architectures considered. 

\section{Data sample construction}
\label{sec:data_sample}

Our models should not only be able to distinguish differences in source angular extent, but furthermore be able to find differences in physical linear sizes. Therefore the data set sample selection in this work needs to build a data set that includes confirmed GRGs and radio galaxies with smaller sizes. The data sample selection should also follow the rules of data set foundation for training Multi-branched CNNs. Consequently, the data sets used in this work were selected in accordance to the following criteria:
\begin{description}
    \item[\textbf{Data availability}:] All data samples should include (i) image survey data from both the NVSS and FIRST surveys, (ii) host galaxy redshifts, (iii) Largest Angular Size (LAS) measurements, and (iv) have a physical linear size calculated.
    \item[\textbf{Source-Image relationship}:] Image data from each radio survey should have the same image size in terms of angular size.
    \item[\textbf{Image Pre-processing}:] After pre-processing images should only contain positive-valued pixels and the source should be visible in the image.
    \item[\textbf{Traceability}:] Users should be able to trace the source coordinates, catalogued object ID, and original source catalogue for each sample.
\end{description}

With respect to data availability, in this work we only include samples that have both NVSS and FIRST data available. \textcolor{black}{We do not require the data availability of LOFAR Two-metre Sky Survey Data Release 1 \citep[LoTSS DR1;][]{2019A&A...622A...2W} conducted by the LOw Frequency ARray \citep[LOFAR;][]{2013A&A...556A...2V}. This is because} LoTSS DR1 covers only 424 square degrees and this would significantly limit data availability for the samples selected in Section~\ref{sec:sample_selection}. Data traceabilty is implemented not only for the convenience of model training and testing, but allows both users and developers to evaluate and explain model outcomes based on their scientific understanding of the data. For example, in this work, we use data traceability in Section~\ref{sec:discussion} to explain misclassifications with respect to different models. 

\subsection{Source sample selection}
\label{sec:sample_selection}

\subsubsection{Radio galaxies of smaller sizes}

Radio galaxies with non-giant dimensions were selected from Data Release 1 of Radio Galaxy Zoo (RGZ DR1; Wong et al. in prep.). RGZ DR1 is a radio galaxy catalogue created by over 12\,000 volunteers through the RGZ citizen science project. Project users are asked to cross match radio source lobes with a corresponiding infrared host galaxy. The radio images come mainly from the FIRST survey, and the infrared images are largely 3.4\,$\mu$m $WISE$ images. These radio and infra-red images share a $3\times3$\,arcmin field of view. RGZ DR1 includes 75\,641 cross-matched identifications (Wong et al. in prep.).

A previous investigation of the DR1 catalogue has shown that the uniform $3\times3$\,arcmin image size constrains its ability to identify GRGs \citep{2020MNRAS.499...68T}. However, the catalogue does provide a large source sample with full data availability and traceability, as required for the construction of a machine learning dataset. Previous analysis of the full RGZ DR1 catalogue found that at least 11\,237 non-duplicated samples fulfill the requirements for training data set selection outlined at the start of Section~\ref{sec:sample_selection}. These samples have LAS from 2 arcsec to 195 arcsec, and are all within the $3\times3$\,arcmin field of view.

In addition to the source-image relationship required for each data set sample, we further require that the radio centroid and host galaxy position should be consistent for each galaxy, as asymmetries in sources positions are known to exist among the RGZ DR1 catalogued samples (Wong et al. in prep.). Consequently, we only retain sources with an angular separation between the host galaxy and estimated radio centre smaller than 1.8\,arcsec, which is the pixel size of FIRST survey images \citep{1995ApJ...450..559B}. This criterion is to ensure sample symmetry and reduces the 11\,237 RGZ DR1 samples to 6\,021 samples. On inspection, we also removed the known GRG source GRG\,J1402+2442 from the sample, as we only consider objects with smaller sizes ($\le$700\,kpc) in the RGZ DR1 sample. Finally, we examined the survey image data availability of NVSS and FIRST using the SkyView API query {\tt astroquery.skyview.get\_image\_list}, to confirm that all sources had the required image data. 

\subsubsection{Giant Radio Galaxies}

The GRG sample for this work comes from the \cite{kuzmicz2018} and \cite{dabhadelotss} catalogues. \cite{kuzmicz2018} performed a detailed review of the literature identifying 349 GRGs, of which 89.7\% are FR\,II objects. The catalogue has its sources validated using the NVSS at 1.4\,GHz and measures their flux densities using those data. However we note that primary image data used for GRG identification in the literature covers a wide range of image angular resolutions, from arcsec level to 45$''$ (e.g. LoTSS, NVSS, SUMSS). 

\cite{dabhadelotss}, on the other hand, performed an independent GRG search using the Value Added Catalogue \citep[VAC;][]{2019A&A...622A...2W} of LOFAR. The team identified 239 GRGs, including 225 which were previously unknown. The newly discovered GRGs in the catalogue of \cite{dabhadelotss} have their candidates identified from LoTSS survey images at 151\,MHz with 6$''$ resolution. They cross-validated their sources using the  FIRST, WENSS and TGSS surveys \citep{dabhadelotss}. The difference in GRG sample selection and cross-validation between these samples will allow us to compare model behaviour when using samples selected with different class definitions. We discuss this further in Section~\ref{sec:model_performance}.

From these catalogues we found that 310 GRGs in the \cite{kuzmicz2018} catalogue have NVSS images available, and 186 also have FIRST images. All newly discovered GRGs in the \cite{dabhadelotss} sample have both NVSS and FIRST images available. Considering that the maximum source LAS in the RGZ DR1 entries is 195\,arcsec, we further require the GRG samples to have LAS equal or smaller than the DR1 LAS limit. This reduces the samples to 58 GRGs from the \cite{kuzmicz2018} catalogue and 167 GRGs from the \cite{dabhadelotss} catalogue.

\subsection{Image pre-processing and further sample selection}
\label{sec:pre_processing}

In this work all image data are obtained using the Skyview Virtual Observatory\footnote{\url{https://skyview.gsfc.nasa.gov}}. Consistent with the RGZ DR1, we define our \textcolor{black}{FIRST} image data to have a field of view with a uniform $3\times3$\,arcmin size. This is equivalent to 100$\times$100\,pixels for the FIRST images where 1\,pixel = 1.8$''$. \textcolor{black}{Since NVSS has an angular resolution of 45$''$, we define 18$\times$18\,pixels dimensions for the NVSS images, where 1\,pixel = 15$''$, in order to avoid truncating objects with large angular extents that have emission close to the image boundaries.} We acquire both the NVSS and FIRST postage stamp images in FITS format, defining the image centres using the host galaxy position from RGZ~DR1. 

The original FITS images are linearly scaled, and have units of Jy/beam. Following the literature, we then subject each FITS image to a series of pre-processing steps before use. \cite{2017ApJS..230...20A} highlighted the importance of image noise reduction, and replaced all image pixels with values lower than a specified noise threshold with zeros, giving the sample images cleaner backgrounds and enabling neural networks to train with high sparsity. Specifically, they proposed this sigma-clipping for each sample image to be at the 3-$\sigma_{\rm rms}$ level. Later studies showed that applying sigma-clipping at a 3, 4 or 5-$\sigma_{\rm rms}$ level did not significantly change the model outcome, and the approach of sigma-clipping generally has been applied successfully by a number of applications \citep[e.g.,][]{2017ApJS..230...20A,2019MNRAS.488.3358T}.

In this work, we also follow the method of \citet{2017ApJS..230...20A} and sigma-clip the images individually at a level of 3$\sigma_{\rm rms}$. We then apply the following image normalization to each of the images:
\begin{ceqn}
\begin{align}
\rm Output = \frac{Input - Min}{Max - Min} \times (255.0 - 0.0).
\label{eqn:I}
\end{align}
\end{ceqn}

Following this pre-processing, we found that some radio sources with very low signal to noise ratios were eliminated. We found that 15, 7, and 15 objects from the DR1, Kuzmicz and Dabhade source catalogues, respectively, had at least one of the FIRST or NVSS images result in an empty field of view and consequently these objects were removed from our sample.

\begin{figure}
\setlength{\unitlength}{1cm}
\includegraphics[width=0.45\textwidth]{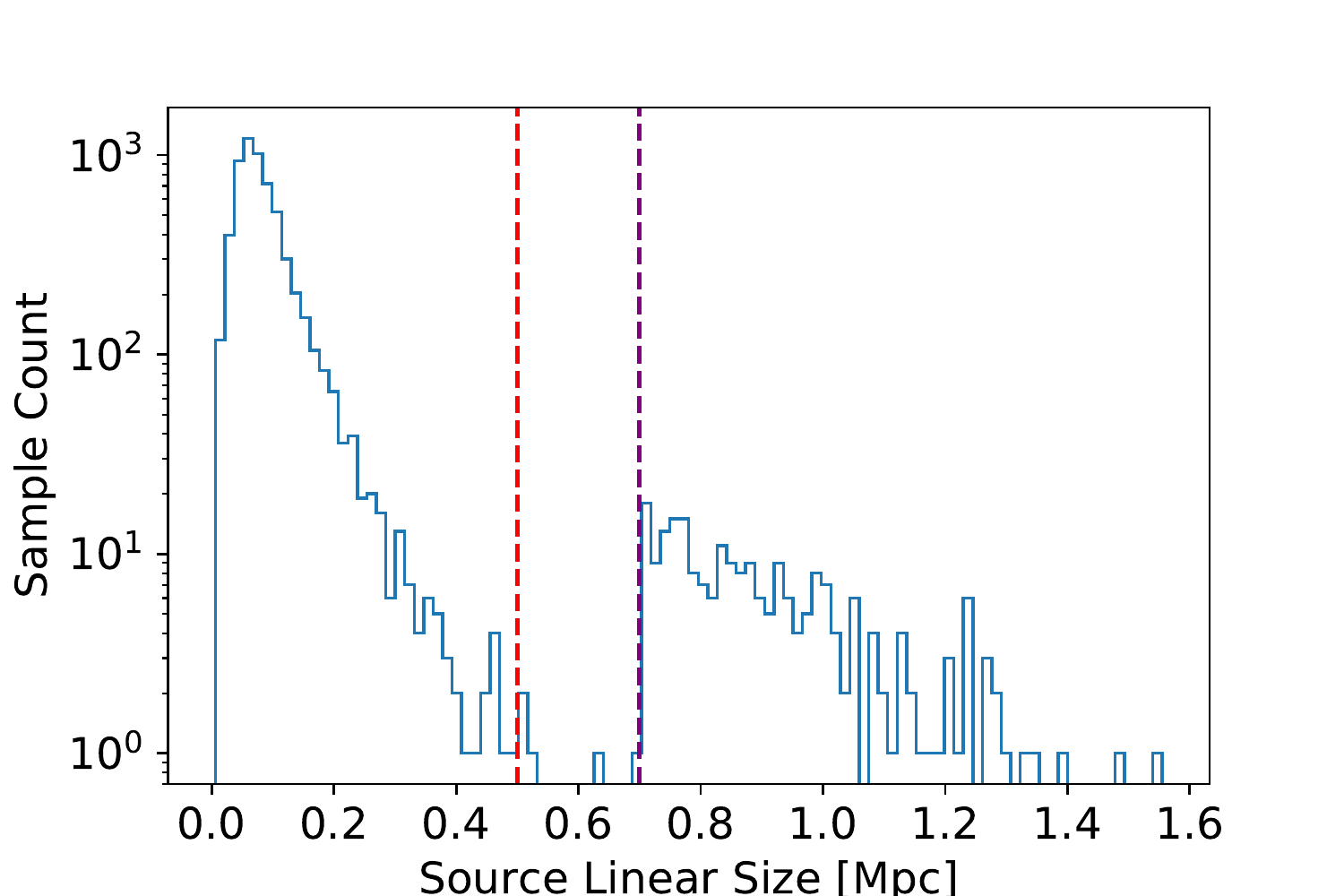}
\caption{Bulk sample projected linear size distribution after we performed sample section procedure in Section~\ref{sec:sample_selection} and Section~\ref{sec:pre_processing}. The red and blue dashed lines on the diagram indicate projected linear sizes of 500\,kpc and 700\,kpc, respectively.}
\label{fig:Sample_size_vs_num}
\end{figure}

Figure~\ref{fig:Sample_size_vs_num} shows the size distribution of the remaining objects, where the red dashed line indicates a source size of 500\,kpc and the blue line represents 700\,kpc, the GRG size cut-off. Only four objects have linear sizes intermediate to these two values and, for clarity, we exclude these four intermediate sources from the data set and define the two target classes of radio galaxy in this work to be:

\begin{itemize}
    \item NOM: Radio galaxies with linear sizes smaller than 500\,kpc.
    \item GRG: Radio galaxies with linear sizes larger than 700\,kpc, consistent with the standard definition from the literature.
\end{itemize}

Following the pre-processing described above, the sample contains 6001 radio galaxies of class NOM, and 205 of class GRG. Since the two classes have a clear difference in linear size, in principle a good classifier should be able to distinguish them well. All of these objects are centred in their images. Data for each sample also includes source object ID, qualified host galaxy redshift, LAS measurement, and computed source linear size. However, as described in Section~\ref{sec:network_architecture}, only image data and redshift information are used as model inputs for the classifiers evaluated in this work.

\subsection{Data formating, division and summary}
\label{sec:data_formating}

Although there are $\sim6000$ sample sources of class NOM, the dataset would be extremely imbalanced if we use all of them when building our training/testing set. In consideration of the observationally imbalanced nature of GRGs and sources with smaller sizes, we use these data to build a modestly imbalanced dataset of 600 training samples and 200 testing samples (3 to 1 ratio), with GRG samples extracted from the \citet{kuzmicz2018} catalogue only, and samples of class NOM taken from RGZ DR1. The resulting dataset is named GRGNOM-A, and has a NOM:GRG class balance of $\sim 14:1$.  Given the availability of the \citet{dabhadelotss} samples, we also build a second dataset named GRGNOM-B using 204 out of 205 GRGs in the selected data samples. The 204 objects would be separated into 153/51 GRG samples in the resulting GRGNOM-B training/testing set. Sources in both class GRG and NOM in the GRGNOM-B dataset then would follow the 3 to 1 data sample ratio. For either GRGNOM-B training set or testing set, source NOM:GRG sample ratio in each subset is around $2.9:1$. Given that the test set of GRGNOM-A only contains \citet{kuzmicz2018} samples, we only include \citet{dabhadelotss} samples in the GRGNOM-B test set in order to allow models trained using either data set to test their generalization ability. The resulting data samples are summarized in Table.~\ref{tab:dataset_division}.   
\begin{table}
\begin{center}
\begin{tabular}{| c | c | c | c | c |}
\hline \hline
\textbf{GRGNOM-A} &  \textbf{NOM} &  \multicolumn{2}{ c |}{\textbf{GRG}} & Total   \\
               &  \textbf{RGZ DR1} & \textbf{Kuzmicz} & \textbf{Dabhade} & \\
\hline
\textbf{Training} & 561   & 39  & 0  & 600  \\ 
\textbf{Testing}  & 187   & 13  & 0  & 200  \\ \hline 
Count             & 748   & 52  & 0  & 800 \\ \hline \hline
\textbf{GRGNOM-B} &  \textbf{NOM} &  \multicolumn{2}{ c |}{\textbf{GRG}} & Total   \\
               &  \textbf{RGZ DR1} & \textbf{Kuzmicz} & \textbf{Dabhade} & \\
\hline
\textbf{Training} & 447   & 52  & 101 & 600  \\ 
\textbf{Testing}  & 149   & 0  & 51  & 200  \\ \hline 
Count             & 596   & 52  & 152 & 800 \\ \hline \hline
\end{tabular}
\end{center}
\caption{A summary of the sample division of the GRGNOM-A and GRGNOM-B. `Count' refers to the total source sample number of a class, and `Total' represents the sample number of each column.}
\label{tab:dataset_division}
\end{table}

It can be seen that the GRGNOM-B training sample is dominated by \citet{dabhadelotss} samples primarily identified at 151\,MHz, while the GRGNOM-A data set only includes \citet{kuzmicz2018} samples identified at 1.4\,GHz. In particular, the test set of GRGNOM-B only contains samples from the \citet{dabhadelotss} catalogue, in order to see if the features learned from \citet{kuzmicz2018} samples can facilitate the identification of \citet{dabhadelotss} samples when looking at their 1.4\,GHz radio image data. The data sample construction also gives those models trained with GRGNOM-B a chance to test their generalization ability upon samples with their identification, source LAS, and host galaxy redshifts measured in a uniform manner.

In terms of data format, both GRGNOM-A and GRGNOM-B are split into two parts: (i) text-format tables of numerical source information, e.g. Table.~\ref{tab:Dataset_manifest_example}, and (ii) a group of machine readable documents containing feature data in various formats. This second component of the data set is comprised of:
{\renewcommand\labelitemi{}
\begin{enumerate}
  
    \item \textbf{FIRST images}: The pre-processed FIRST survey greyscale images with a universal size of 100 $\times$ 100 pixels. Three versions of these image files are generated:
        \begin{itemize}
        \item \textbf{a. Image batch files}: Images are saved in 4 batched files (3 training batches and 1 testing batch), along with a metadata file containing corresponding image header information. These files are in a format that is understandable for our {\tt Pytorch} models. These files are saved in a folder named FIRST.
        
        \item \textbf{b. Encoded compressed file}: This is the encoded compressed file of \textbf{a}. The compressed file is named as FIRST.tar.gz. Creating such a data file allows future developers to download and re-use the image data samples for machine learning training.
        
        \item \textbf{c. Individual images}: These images are saved in another image folder named {\tt FIRST$\_$IMG}. Image names follow the format: "Catalogue Name$\_$Catalogue Source No$\_$Right Ascension$\_$Declination.png".
        \end{itemize}
        
    \item \textbf{NVSS images}: The pre-processed NVSS survey greyscale images with a universal size of 18 $\times$ 18 pixels are saved in the same manner as the FIRST images.
    
    \item \textbf{Source host galaxy redshift}: Consistent with the image data, numerical source host galaxy redshifts are separated into 4 batches and saved as {\tt numpy} arrays.
    
    \item \textbf{Source LAS}: Numerical source LAS are organized and saved in the same way as (iii).
    
    \item \textbf{Source linear size}: Numerical source linear size are organized and saved in the same way as (iii).
    
    \item \textbf{Source object ID}: Primary object ID of each source sample in their original catalogues. These data strings have been encoded in the {\tt uint8} format and saved in the same way as (iii). As necessary, they can be decoded in the {\tt utf-8} format.
    
    \item \textbf{Class label}: Numerical class label of each data sample: 0 and 1 represent source classes NOM and GRG, respectively. They are organized and saved in the same way as (iii).  

\end{enumerate}
}

By using the Python {\tt pickle} package, we built our training/testing datasets with these components, which allow users to call any of the source data samples in the dataset. Hash values were generated separately for (i)a, (i)b, (ii)a and (ii)b to protect their integrity and avoid manipulation. 

We note that although source LAS and physical size are included in the data set for completeness, this information is not used to train any of the models in this work.

\begin{table*}
\begin{center}
\begin{tabular}{| p{2.5cm} | p{2cm} | p{2cm} | p{1.5cm} | p{1.5cm} | p{1.5cm} | p{1.5cm} |}
\hline

 \textbf{Object ID} & \textbf{RA (J2000.0)}  & \textbf{DEC (J2000.0)}  & \textbf{z} & \textbf{LAS} & \textbf{Size} & \textbf{Label}\\
   &  [h:m:s]  & [d:m:s]  &  & [arcsec] & [Mpc] & \\
\hline
RGZJ000606.0+013125 &	00:06:06.07&	01:31:25.20&	0.23372&	21&	0.079&	NOM\\
RGZJ000626.4+081838 &	00:06:26.41&	08:18:38.49&	0.41540&	18&	0.102&	NOM\\
RGZJ000627.2+060407 &	00:06:27.21&	06:04:07.29&	0.30091&	14&	0.064&	NOM\\
RGZJ000746.4+031938 &	00:07:46.46&	03:19:38.99&	0.29194&	44&	0.200&	NOM\\
RGZJ000851.0+045243 &	00:08:51.01&	04:52:43.92&	0.34255&	19&	0.095&	NOM\\
RGZJ000911.0+145105 &	00:09:11.05&	14:51:05.07&	0.36832&	20&	0.105&	NOM\\
RGZJ001042.9+091917 &	00:10:42.92&	09:19:17.44&	0.15308&	19&	0.052&	NOM\\
RGZJ001051.0+141655	&   00:10:51.09&	14:16:55.86&	0.31507&	17&	0.079&	NOM\\
RGZJ001146.7+101528 &	00:11:46.75&	10:15:28.45&	0.22175&	17&	0.064&	NOM\\
RGZJ001524.2+143038 &	00:15:24.23&	14:30:38.83&	0.22668&	33&	0.122&	NOM\\
\hline 
\end{tabular}
\end{center}
\caption{The first 10 rows of the GRGNOM-A training sample catalogue. Source object ID, RA/DEC, host galaxy redshift ($z$) and LAS are extracted from RGZ DR1, while the source linear size is derived from the $z$ and LAS of each sample based on the  cosmological parameters defined in \citet{2016A&A...594A..13P}. Class labels are defined as described in Section~\ref{sec:pre_processing}.}
\label{tab:Dataset_manifest_example}
\end{table*}

\begin{figure}
    \setlength{\unitlength}{1cm}
    \centering
    \includegraphics[width=0.48\textwidth]{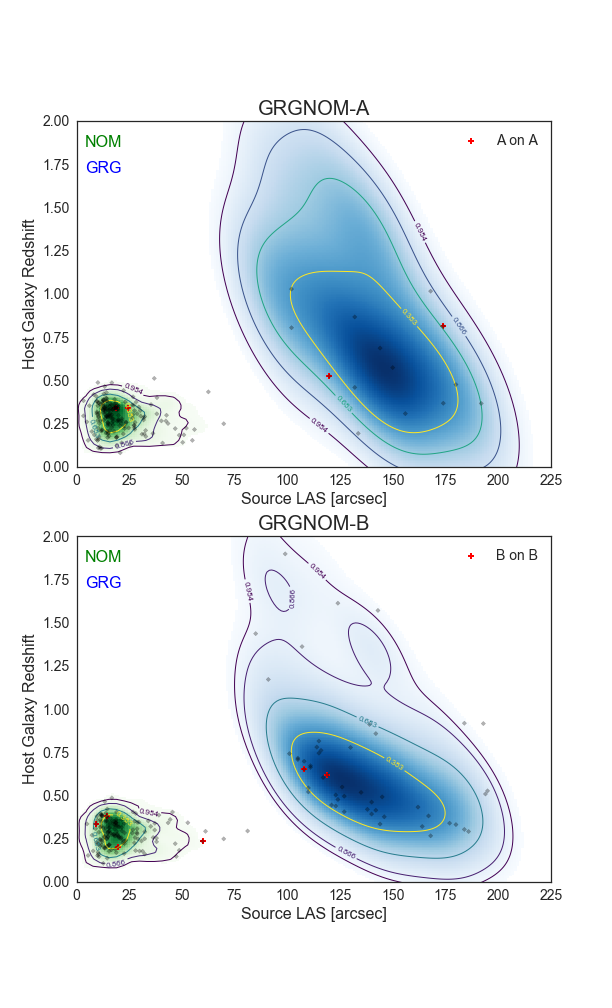}
    \caption{Upper: the LAS vs. host galaxy redshift density map of the GRGNOM-A dataset. Training samples of class NOM and GRG are represented in green and blue, respectively. Contours on the diagram refer to the iso-proportion of the density, i.e. 95.4\% of the probability mass lies within the 0.954 contour. Contours are shown at 0.383, 0.683, 0.866 and 0.954 in the figure, which correspond to 0.5, 1, 1.5 and 2$\sigma$. Grey data points are the GRGNOM-A test sample data points and red data points on the diagram indicate samples that are frequently mis-classified by models of Architecture~G trained and tested on the GRGNOM-A data set. Lower: equivalent distributions for the GRGNOM-B dataset. The red data points indicate samples that are frequently mis-classified using Architecture~G trained and tested on the GRGNOM-B data set. This diagram was plotted using the {\tt pyrolite} \citep{Williams2020} package. \label{fig:GRGNOM_parameter}}
\end{figure}

\subsection{Data Normalization and Augmentation}

\begin{figure*}
    \centering
    \includegraphics[width=\textwidth]{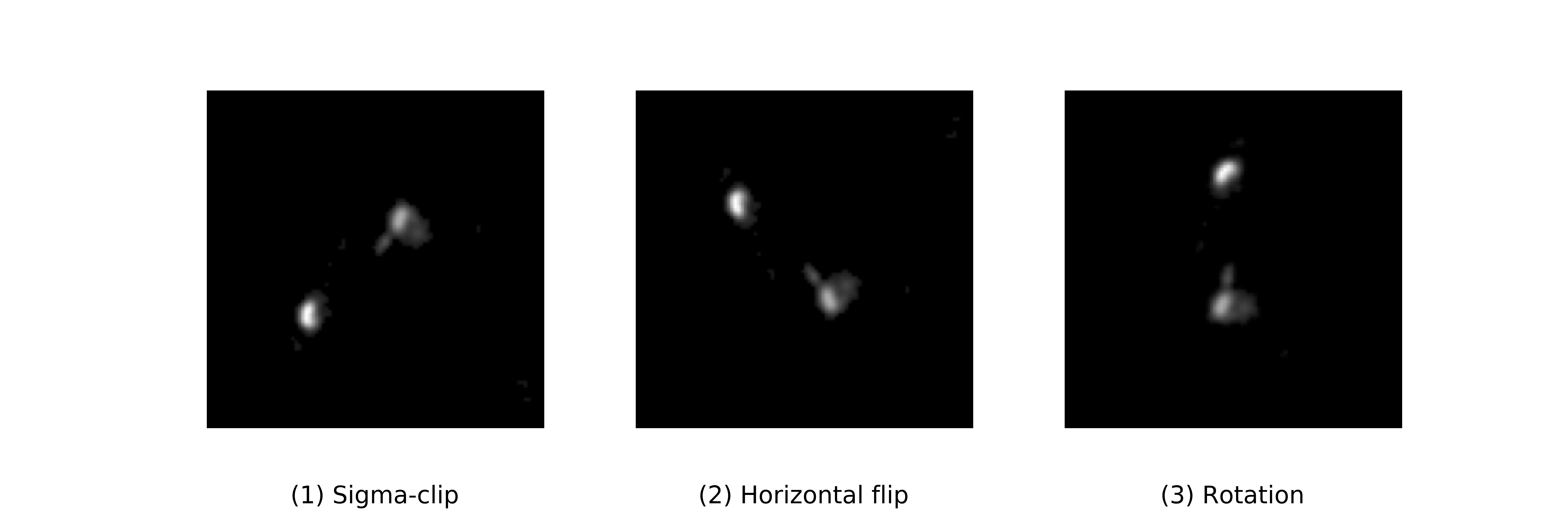}
    \caption{An illustration of image pre-processing and data augmentation using an example FIRST survey image (object id: Dabhade201), which is a radio source of class GRG with LAS of 108\,arcsec.}
    \label{fig:pre_process_steps}
\end{figure*}

In order to improve the convergence of model training, it is recommended to define a data normalization and augmentation strategy \citep{tricks}. Normalization constrains data values within a given range, reduces skew and therefore speeds up the training process of a model. In this work, we require the image data to be normalized to have both a mean and standard deviation of 0.5 before importing to a model, which constrains image pixel values largely within the range from 0 to 1. 

In the case of data augmentation, we apply horizontal flipping and image rotation in this work (i.e.Figure~\ref{fig:pre_process_steps}). Specifically, we perform horizontal flipping of every image sample with a random probability of 50\%. We then further rotate each image in a clockwise manner using a randomly selected angle from \textcolor{black}{-180$^{\circ}$ to 180$^{\circ}$}, where the rotated angle should be an integer in units of degrees. 

Data augmentation is performed dynamically during training as the data are imported to the model. This strategy both ensures the model has a large enough number of data samples to learn and minimises memory usage. \textcolor{black}{Using this approach, the augmented training data set has a size of 600 $\times$ 2 $\times$ 360 = 432\,000 samples when the model is trained for 720 epochs.}

\section{Network Architecture}
\label{sec:network_architecture}

In this work we consider five different network architectures to create seven different models. These are summarised in Table~\ref{tab:architectures_summary}. The first network architecture is a traditional, or classical, CNN approach that takes a single source of image data as an input, this forms the basis for Architecture~A (NVSS) and Architecture~B (FIRST) in Table~\ref{tab:architectures_summary}. The second form of network is a multi-domain architecture that takes a single source of image data plus redshift information as its inputs, this forms the basis for Architecture~C (NVSS + $z$) and Architecture~D (FIRST + $z$) in Table~\ref{tab:architectures_summary}. The third form of network is a multi-domain network that takes multiple sources of image data as inputs, this forms the basis for Architecture~E (NVSS + FIRST), and the fourth form of network is the expansion of this architecture to include redshift as an input, Architecture~F (NVSS + FIRST + $z$). The final form of network is Architecture~G in Table~\ref{tab:architectures_summary}, which has the same inputs as Architecture~F but replaces all convolutional layers with Inception Modules, see Section~\ref{sec:inception_module}.

\subsection{Classical CNN}
\label{sec:CNN_logistic_regression}

\cite{2014ApJ...781..117Z} were able to train their pulsar identification algorithms with 3,756 labelled 48 $\times$ 48 images samples using a slightly modified LeNet-5 CNN architecture. LeNet-5 is one of the earliest Convolutional Neural Networks, demonstrating high success in digit/character recognition tasks \citep{Lecun98}. The network has a simple 7-layer architecture, with 2 convolutional layers, 2 pooling layers, and 3 fully-connected layers. In this work, we start with a modified version of LeNet-5, including one extra convolutional layer, see Table~\ref{tab:lenet-5_architecture}. This extra convolutional layer is followed by a down-sampling that differs between the FIRST and NVSS survey images: FIRST images are downsampled to 25 $\times$ 25 = 625, while NVSS images are downsampled to 9 $\times$ 9 = 81.  Rather than using the Mean Squared Error (MSE) originally proposed for LeNet-5, we use the now more common cross-entropy loss function to train our logistic regression algorithms. The layers shown in Table~\ref{tab:lenet-5_architecture} form the base architecture of all networks used in this work and we will refer to specific layer numbers from Table~\ref{tab:lenet-5_architecture} whenever we manipulate or replace any functionality in the following sections.

Although such a network is sufficient to train a model, previous deep learning attempts at classifying radio galaxy morphology have applied additional regularization methods in order to improve their model generalization (test) error and avoid model over-fitting \citep{GoodBengCour16}. In this work, we apply the independent component (IC) layer regularization strategy of \cite{2019arXiv190505928C} to all convolutional layers in our network and this is described in more detail in the following section. In addition, \textcolor{black}{we include a dropout layer before each fully-connected layer, with the exception of the output layer. We note that AlexNet \citep{NIPS2012_4824}, a well-known CNN architecture, has previously been used to classify radio galaxy morphologies \citep{NIPS2012_4824,2017ApJS..230...20A}. Following hyper-parameter optimization, we found that AlexNet requires 3-4 times the training time per epoch compared to the modified LeNet-5 used in this work, and does not provide comparable or improved model performance. }

\setlength{\tabcolsep}{4pt}
\begin{table*}
    \begin{center}
    \begin{tabular}{c|c|c|c|c|c|c|c}
    \hline
    Layer No.     & Layer Type   &  Input Channels & Output Channels & Kernel Size & Stride &  Activation & Regularization \\ \hline
    1  & Convolutional   & 1  &  6  &  5  &  1  &ReLU & \textbf{IC}/\textbf{BN} \\
    2  & Max Pooling & 6  &  6  &  2  &  2  &\\
    3  & Convolutional   & 6  &  16 &  5  &  1  &ReLU & \textbf{IC}/\textbf{BN} \\
    4  & Max Pooling & 16 &  16 &  2  &  2  &\\
    5  & Convolutional   & 16 & 120 &  5  &  1  &ReLU & \textbf{IC}/\textbf{BN} \\
    5$'$ & Squeeze layer 5 outputs  &  &  &    &    &  &  \\
    6  & Fully-connected & 120$\times$Down-sampled neuron number & 120  &     &     &ReLU & Dropout\\
    7  & Fully-connected & 120& 84  &     &     &ReLU & Dropout \\
    8  & Fully-connected & 84 & 2  &     &     &Softmax\\ \hline
    \end{tabular}
    \end{center}
    \caption{A summary of the modified LeNet-5 architecture used in this work as a base architecture. \textcolor{black}{IC refers to the independent component \citep{2019arXiv190505928C} and BN represents batch normalization. 
    }}
    \label{tab:lenet-5_architecture}
\end{table*}

\begin{table*}
    \begin{center}
    \begin{tabular}{l|c|c|c|c|c|c|c}
    \hline
    Architecture     & A   &  B & C & D & E & F & G \\ \hline
    Input data  & \textbf{NVSS} &  \textbf{FIRST}  &  \textbf{NVSS} \& \textbf{z}  &  \textbf{FIRST} \& \textbf{z}  &  \textbf{NVSS} \& \textbf{FIRST}  & \textbf{NVSS} \& \textbf{FIRST} \& \textbf{z} & \textbf{NVSS} \& \textbf{FIRST} \& \textbf{z} \\
    Convolution Branches  & 1   & 1 & 1 &  1  &  2  & 2 & 2\\
    Layer 6 input (\textbf{NVSS}) & 120 $\times$ 9 $\times$ 9 &   &  120 $\times$ 9 $\times$ 9  &   &  120 $\times$ 9 $\times$ 9 & 120 $\times$ 9 $\times$ 9  & 128 $\times$ 9 $\times$ 9\\
    Layer 6 input (\textbf{FIRST})  &    & 120 $\times$ 25 $\times$ 25  &   &  120 $\times$ 25 $\times$ 25  &  120 $\times$ 25 $\times$ 25  & 120 $\times$ 25 $\times$ 25 & 128 $\times$ 25 $\times$ 25\\
    Layer 7 input  & 120 & 120 &  120 + 1 &  120 + 1  &  120 + 120  & 120 + 120 +1 & 120 + 120 + 1 \\
    Extra FC  & No & No &  No & No  &  Yes  & Yes & Yes \\
    Branch Module  & Nil & Nil & Nil  & Nil & Nil & Nil & Inception Modules \\ \hline
    \end{tabular}
    \end{center}
    \caption{The summary of architectures we adopted in this work. \emph{Convolution Branches} refers to the number of independent top-down architectures from Layer~1 to Layer~5$'$ in Table.~\ref{tab:lenet-5_architecture}.}
    \label{tab:architectures_summary}
\end{table*}

\begin{figure*}
    \centerline{
        \includegraphics[width=0.35\textwidth]{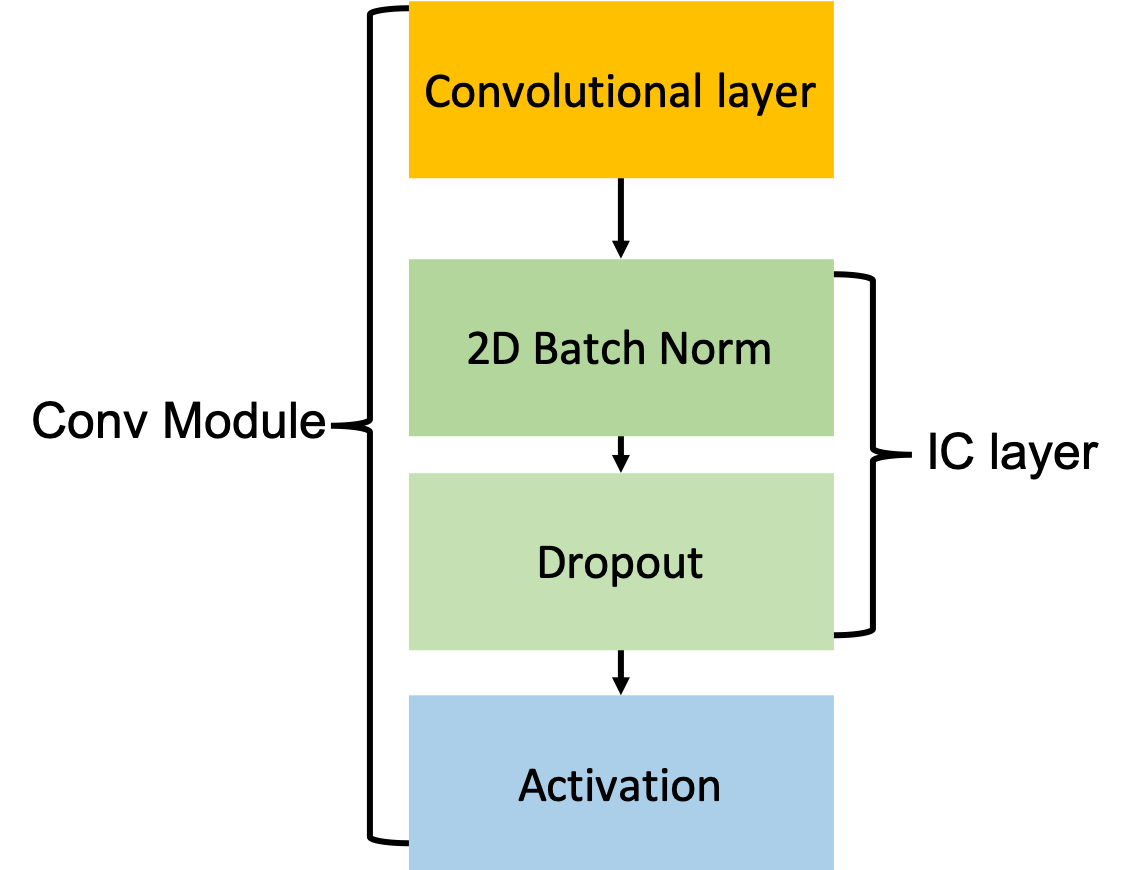}\qquad
        \includegraphics[width=0.55\textwidth]{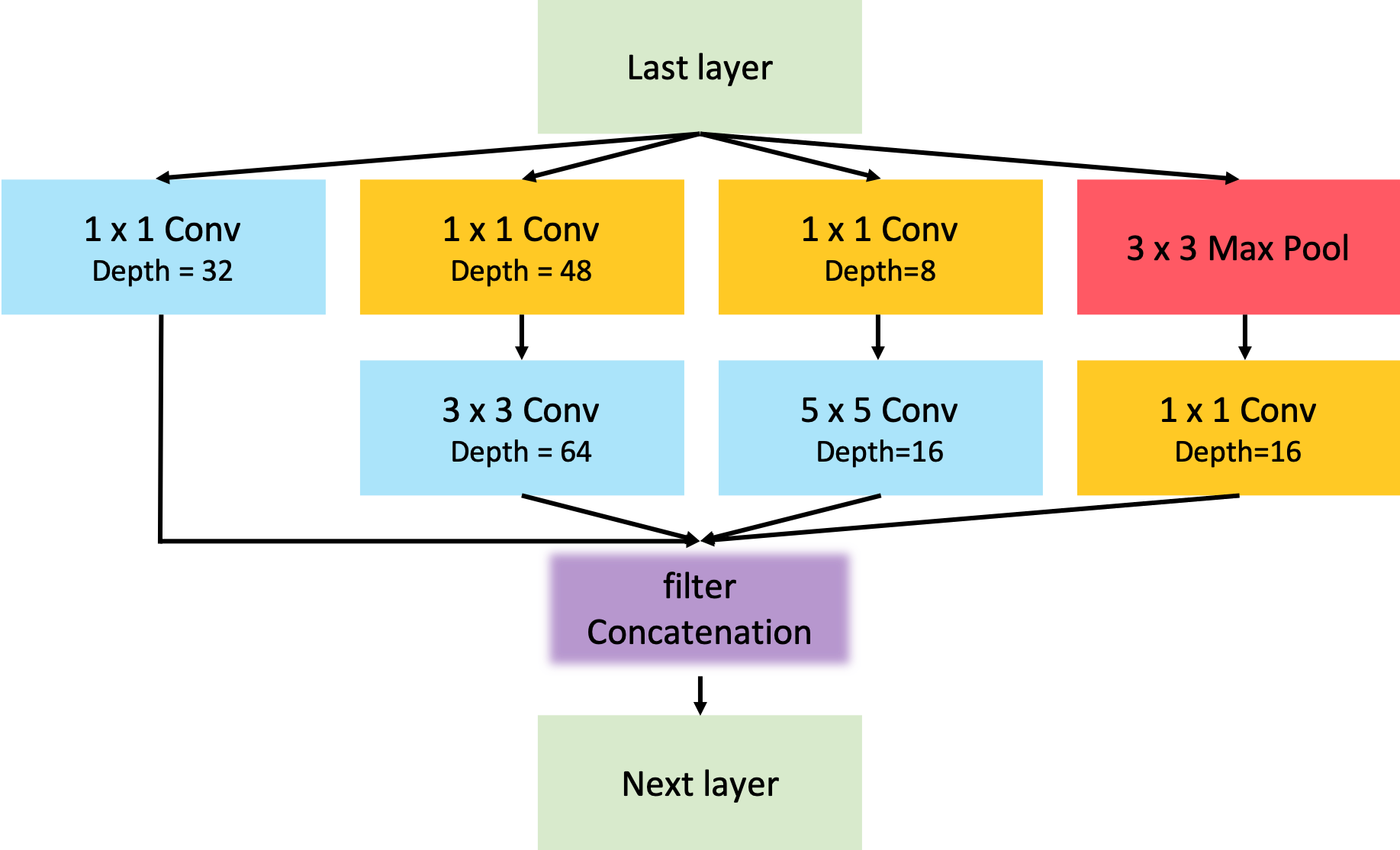}\qquad}
    \caption[]{Left: The Conv module we used in this work, which is inspired by \citet{2018arXiv180105134L}. The \emph{IC layer} refers to the Independent Component layer \citep{2019arXiv190505928C}; Right: The Inception module with dimension reduction \citet{2014arXiv1409.4842S}.}
    \label{fig:module_setup}
\end{figure*}

\subsection{Independent Component Layer}
\label{sec:ic}

As deep learning develops, one important issue is how to train complex networks with higher efficiency \citep{2015arXiv150203167I,2019arXiv190505928C}. Among all the techniques available, Batch Normalization \citep[BN;][]{2015arXiv150203167I} and Dropout \citep{Srivastava2014} are perhaps most frequently used by radio galaxy related deep learning approaches \citep[e.g.][]{2017ApJS..230...20A,2019ApJS..240...34M}.

Batch normalisation is able to normalize the net activations and have their mean and unit variance become zero \citep{2019arXiv190505928C}. The purpose of applying such an approach is to reduce the internal covariate shift, in other words the change in the distribution of network activations due to the change in the network parameters during training \citep{2015arXiv150203167I}. The technique therefore is able to speed up network training, regularize model performance, and further induce a stable predictable behaviour during gradient descent \citep{2018arXiv180511604S}.

Dropout, on the other hand, performs regularization in a different way. It introduces random gates for all inputs to a given layer, where each neuron has a probability, $p$, to be set to zero. Such a  measure is able to remove weakly connected neurons, and has been demonstrated to regularize network performance and prevent neuron co-adaptation \citep{Srivastava2014}.

The Independent Component layer (IC) is a recently developed technique incorporating both of these techniques that has been proposed to boost model training efficiency and improve model stability \citep{2018arXiv180105134L,2019arXiv190505928C}. Each IC layer contains a stacked combination of BN and Dropout layers, see Figure~\ref{fig:module_setup}, which has been proven to be able to reduce the mutual information and the degree of correlation between any pair of neurons. Such techniques achieve more stable training behaviour and IC networks typically have their generalization ability improved \citep{2019arXiv190505928C}. A recent approach put IC layers before the activation layers \citep{2018arXiv180105134L}, and found that such method could boost model performance compared with those which only insert a BN layer between a convolutional layer and an activation function.

Inspired by \citet{2018arXiv180105134L}, in this work we replace each convolutional layer with the combination of a convolutional layer, an IC layer and an activation function and refer to this as a \emph{Conv module}. This is illustrated in Figure~\ref{fig:module_setup}. We did not use the \citet{2019arXiv190505928C} strategy as to maintain image input completeness. \textcolor{black}{We also implemented the popular Conv->BN->ReLU regularization strategy \citep[e.g.,][]{2015arXiv151203385H,2019MNRAS.488.3358T} for further comparison.}

\subsection{Inception Module}
\label{sec:inception_module}

The \emph{Inception module} was created by \citet{2014arXiv1409.4842S} and is named following a previous approach referred to as Network in Network \citep[NIN;][]{2013arXiv1312.4400L}. Unlike classical CNN approaches that have all of their convolutional layers stacked sequentially, the Inception module has 4 `branches': a convolutional layer with limited kernel sizes of 1 $\times$ 1, 3 $\times$ 3, 5 $\times$ 5 and a 3 $\times$ 3 max pooling layer. The 4 branches in each module operate in parallel on the same input feature map. Outputs from each `branch' then are concatenated together and serve as the input to the next layer, see Figure~\ref{fig:module_setup}. 

In order to give the network improved representation power, \citet{2014arXiv1409.4842S} further introduced a 1 $\times$ 1 convolutional layer both before the 3 $\times$ 3 and 5 $\times$ 5 convolutional layers, and after the max pooling layer. An Inception module with these additional layers is known as an \emph{Inception module with dimension reduction} \citep{2014arXiv1409.4842S}. The addition of these 1 $\times$ 1 convolutional layers can reduce the number of input channels, hence lowering the computational complexity by serving as dimension reduction module, whilst increasing network depth at the same time. 

\textcolor{black}{The first architecture equipped with the Inception module with dimension reduction was GoogLeNet, the winner of the ImageNet Large-Scale Visual Recognition Challenge 2014 \citep[ILSVRC14;][]{2014arXiv1409.0575R}. After the success of GoogLeNet, this module was also used} in astronomical studies such as supernovae classification \citep{2019arXiv190100461B} and Faraday spectra classification \citep{2019MNRAS.483..964B}, and the mapping between simulation based galaxy cluster distributions and the underlying dark matter distribution \citep{2019arXiv190205965Z}. \textcolor{black}{In this work, we explore its potential when creating Architecture~G, see Table~\ref{tab:architectures_summary} and Section~\ref{sec:Multi_branch_CNN}}.

\subsection{Multi-domain CNNs}
\label{sec:Multi_domain_CNN}

Although sources with class NOM and GRG are clearly separated in physical linear size, see Figure~\ref{fig:Sample_size_vs_num}, their LAS and host galaxy redshift distributions are not as distinct, see Figure~\ref{fig:GRGNOM_parameter}. Typically, the source LAS of the two classes are generally separated, while it can be seen that host galaxy redshift distribution of the two classes are significantly overlapped. Moreover, a source might exhibit different emission structure between its NVSS and FIRST images due to structure appearing on a range of scales. To account for these possibilities, we modify the original network architecture, allowing multiple inputs to train together. Architectures with multiple inputs are also known as \emph{multi-domain} neural networks, and were originally proposed by \citet{2017arXiv170601788A}. They introduced both spatial domain data (2 dimensional) and frequency domain data (1 dimensional) as network inputs, with each domain training on an independent branch with their outputs concatenated into a single fully-connected layer. For such a multi-domain network, model inputs can be 2D images, 1D arrays, or scalars. Networks with multiple inputs are sometimes considered to be a variant of multi-branch networks, and we discuss this further in Section~\ref{sec:Multi_branch_CNN}.

\begin{figure*}
\setlength{\unitlength}{1cm}
\includegraphics[width=0.9\textwidth,height=0.8\textwidth]{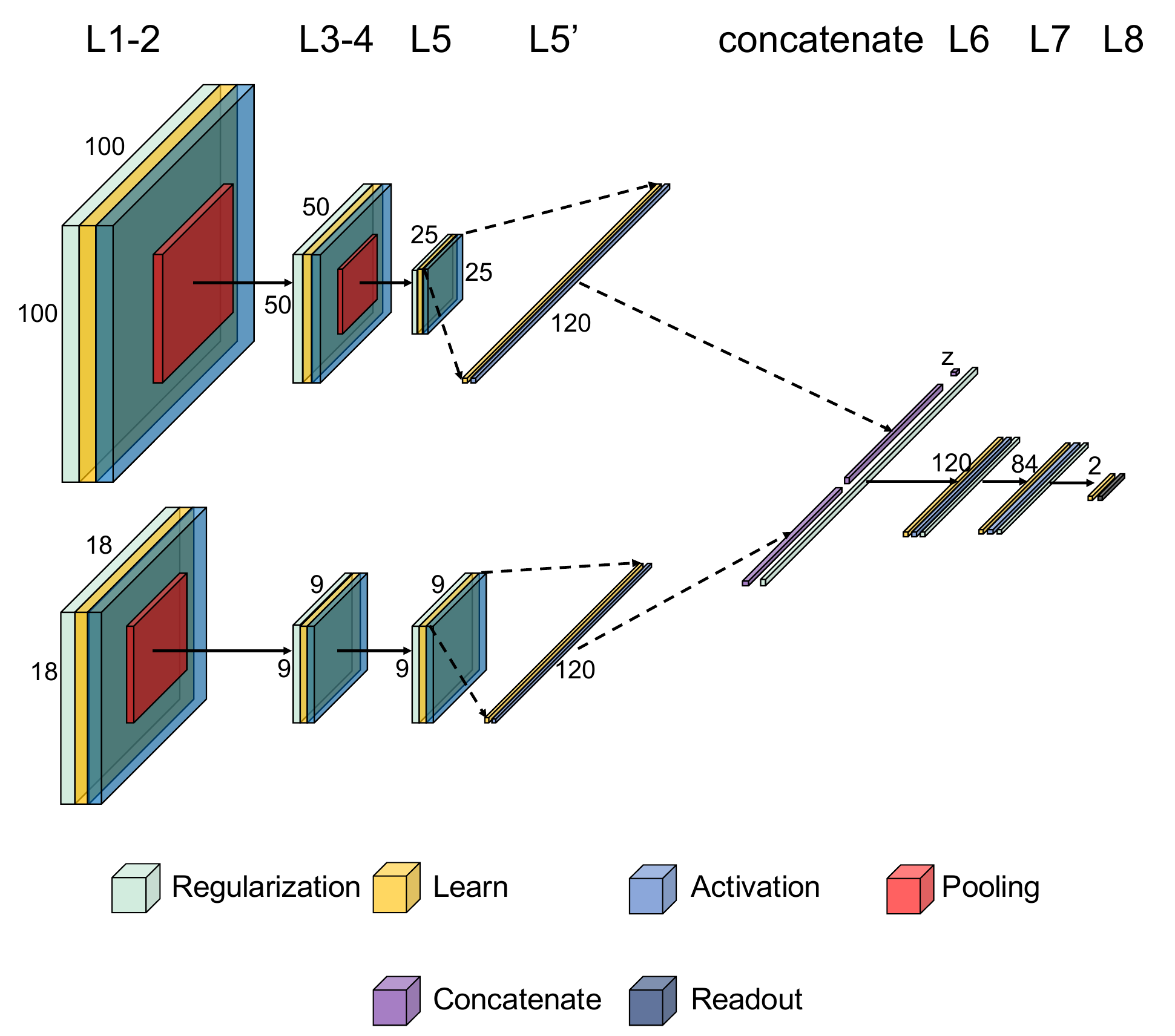}
\caption{The network illustration of the architecture \textbf{F} in our work. In this diagram, {\tt Regularization} refers to the use of \textcolor{black}{IC/BN }layer for the convolutional layers (L1,L3 and L5), and dropout layer for fully-connected layers. {\tt Learn} layers include convolutional layers (L1, L3 and L5) and fully-connected layers (L6-8). {\tt Activation} layers are all ReLU, while {\tt Pooling} layers in the diagram are max-pooling layers. {\tt Concatenate} operation implies that outputs from the last layer and the extra imported parameter (host galaxy redshift) would be concatenated as an 1-D vector input for the next fully-connected layer (L6). Finally, the {\tt Readout} layer is where softmax fucntion is operated, which provides the model class probability prediction. For full parametric details of this architecture, see Table~\ref{tab:lenet-5_architecture} and Table~\ref{tab:architectures_summary}.}
\label{fig:model_architecture_f}
\end{figure*}

\subsubsection{Including source redshift}
\label{sec:Parameter_addition}

Source redshift, $z$, as a numerical parameter, is one of the two key parameters used when identifying GRGs, as it determines the distance to the radio galaxy and hence the conversion of projected angular size to projected physical size. It is therefore intuitive to include source redshift into our models. However, given that numerical data cannot be passed through layers expecting 2-dimensional inputs, we specify for this parameter to join the training at Layer~7,  see Table~\ref{tab:lenet-5_architecture}. Specifically, we concatenate the down-sampled outputs from Layer~6 of the network with the source redshift, an example of which is shown in Figure~\ref{fig:model_architecture_f}. Source redshifts are normalized to lie in the range 0 to 1, consistent with the normalisation of image feature data. This normalisation is discussed further in Section~\ref{sec:d_A}.

\subsubsection{Multiple Image inputs}
\label{sec:multiple_image_inputs}

In addition to combining image data with numerical features such as redshift, we also expand the multi-domain approach further to include additional image inputs. Using this approach, images of radio galaxies observed by multiple surveys with different angular resolutions are able to be learned together. Such an approach has previously been considered with in the field of neuroscience \citep{2018arXiv181102942A}. 

In order to implement this strategy, we combine elements of the two CNNs described in the previous section together, see Figure~\ref{fig:model_architecture_f}. The depth of linear neuron concatenation remains the same as in the \emph{Image + z} methods. By using the resulting architecture, both NVSS and FIRST images of a single source will have 120 features extracted from each survey that are then concatenated and passed into the final two fully-connected layers. In this scenario we add an additional fully-connected layer with size 120 after Layer~6, in order to increase the learning ability of the network given the larger volume of input data as well as to regularize the algorithm further.

With this architecture it is trivial to also include source redshift, $z$, see  Figure~\ref{fig:model_architecture_f}. By adding the source redshift in the same way as in other network architectures, Layer~6 of the network has 241 neurons, while other layers remain unchanged.

\subsection{Multi-branched CNN}
\label{sec:Multi_branch_CNN}

The model architectures we have described so far are top-down architectures, where the input to each layer comes from the previous layer's outputs. Such architectures require their convolutional layers to have a customized kernel and stride size, which can be restrictive when one wants to simultaneously learn general features with different kernel sizes. It is constraints such as these that motivated the invention of CNNs with branched modules and later Multi-branch CNNs \citep[e.g.,][]{2017arXiv171005477L,2020NatSR..10.9486G}.

The very early and well-known branched CNNs include the GoogLeNet and later Inception networks \citep[e.g.,][]{2013arXiv1312.4400L,2014arXiv1409.4842S}. These networks implemented the Inception module structure described in Section~\ref{sec:inception_module}. 

Although the Inception module can be beneficial in terms of model performance, such an architecture can lead to large scale outputs, requiring heavy computation power. In this work, in order to minimise training costs, we adopt the modified Inception Module and use it to replace the final convolutional layers  in Architecture~F, denoted L5 in Figure~\ref{fig:model_architecture_f}, and thus enable those layers to learn features with diverse kernel sizes. The filter dimensions for each layer of the Inception module in this work are chosen to be half of the equivalent value for the `inception (3a)' model of GoogLeNet \citep[e.g.,][]{2014arXiv1409.4842S}, making the output parameter number comparable to that of Layer~6 in Architecture~F. \textcolor{black}{We refer to this modified architecture Architecture~G.}

\subsection{Model training}
\label{sec:training}

\textcolor{black}{In order to simplify model performance comparisons, all model training performed in this work uses the Stochastic Gradient Descent optimizer \citep[SGD;][]{robbins1951} for optimization. Training sample data is imported to each model using a batch size of \textcolor{black}{20}, and training data are shuffled in every training epoch. For model hyper-parameter selection, we perform hyper-parameter grid searches for model architectures A, B and E \textcolor{black}{using the Independent Component layer regularization strategy (see Section \ref{sec:ic}),} along with different initial learning rates (1e-3, 1e-4, 1e-5), dropout rates (0.4, 0.5, 0.6) and training epoch numbers (180, 360, 720, 1080) separately for both the GRGNOM-A and GRGNOM-B datasets.} 

\textcolor{black}{In order to both prevent the resulting models from over-fitting, and to achieve optimal model performances, we compared the learning curves, model AUC values and GRG class recall of the models with different hyper-parameters. By requiring the GRG class recall of a model to be larger than 0.5 and \textcolor{black}{that the model training and validation losses drop together till the end of model training} regardless of architecture (A, B and E), we found that models trained and tested with the GRGNOM-A dataset matched the requirements only when their initial learning rate was 1e-3, the dropout rate was 0.4 and the number of training epoch was equal to 1080. }

\textcolor{black}{We then applied the same requirements to those models trained and tested with the GRGNOM-B dataset, and found that the models matched the criteria when their dropout rate was 0.4, the initial learning rate was 1e-3 and the number of training epochs was no larger than 360. The validation loss comparison between the models trained for 180 and 360 epochs, see Figure~\ref{fig:learning_curve_illustration_GRGNOM_B}, shows that models generally perform better when they are trained for 360 epochs, with an average validation loss lower compared to those trained for 180 epochs. We therefore decided to train the models of all architectures following the aforementioned result, see Table~\ref{tab:hp_summary}. In the rest of this section, we describe the components of these networks in more detail.}

\begin{table}
\begin{center}
\begin{tabular}{| c | c | c |}
\hline \hline
\textbf{Hyper-parameters} &  \textbf{GRGNOM-A} &  \textbf{GRGNOM-B}  \\
\hline
\textbf{Initialization} & From scratch & From scratch \\ 
\textbf{Batch size} & 20   & 20  \\ 
\textbf{Epoch}  & 1080   & 360   \\
\textbf{Learning rate} & 1e-3   & 1e-3 \\
\textbf{Dropout rate}  & 0.4   & 0.4 \\ \hline \hline
\end{tabular}
\end{center}
\caption{A summary of hyper-parameters we used in model training and evaluation. These hyper-parameters are applied to all architectures we used in this work.} 
\label{tab:hp_summary}
\end{table}

\section{Results \& Discussion}
\label{sec:discussion}

\subsection{Model Evaluation Metrics}
\label{sec:model_metrics_general}

In the context of deep learning classification algorithms, popular model evaluation metrics include Accuracy, Recall, Precision, $\rm F_{1}$ score and AUC score, see e.g. definitions in Appendix~A of \citet{2021MNRAS.501.4579B}. These metrics have been widely used previously in the literature to evaluate CNN based radio galaxy classifiers \citep[e.g.,][]{2017ApJS..230...20A,2019ApJS..240...34M,2019MNRAS.488.3358T,2021MNRAS.501.4579B}. 

Performance metrics for the data set as a whole (Accuracy, AUC) and class-specific performance metrics for the GRG class (Precision, Recall) for each of the models considered in this work evaluated against the GRGNOM-A data set are listed in Table~\ref{tab:GRGNOM_A_model_performance_summary_no_dabhade} and against the GRGNOM-B data set in Table~\ref{tab:GRGNOM_B_model_performance_summary_no_kuzmicz}.

\subsection{Model Performance}
\label{sec:model_performance}

\subsubsection{Models trained with GRGNOM-A}
\label{sec:models_trained_with_GRGNOM_A}

\begin{figure*}
\setlength{\unitlength}{1cm}
\includegraphics[width=\textwidth]{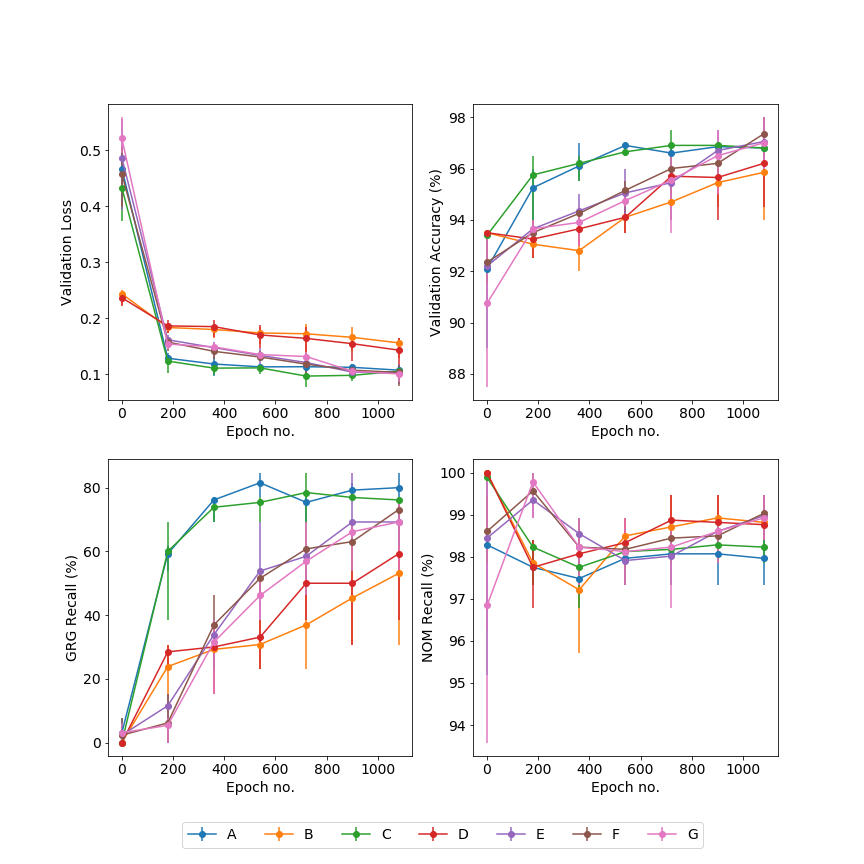}
\caption{The averaged learning curve of the model architectures trained and validated with GRGNOM-A. Assuming normal distribution, the asymmetric errors of each data point on the diagram has covered 60\% of data distribution.}
\label{fig:learning_curve_illustration_GRGNOM_A}
\end{figure*}

Given that the GRGNOM-A data set has a severe class imbalance, class predictions are expected to be biased in the early phases of model training. This can be seen in Figure~\ref{fig:learning_curve_illustration_GRGNOM_A}, where the models used in this work tend to predict almost all validation samples as class NOM in the first \textcolor{black}{200-400} training epochs, although the models do gradually overcome this issue as the training continues. 

Looking at the model loss curves that used the testing set as validation set, it can be seen from Figure~\ref{fig:learning_curve_illustration_GRGNOM_A} that models trained with architectures \textcolor{black}{A and C} have their validation loss saturated quickly, while their ability to classify GRG objects increases gradually as training continues. On the other hand, the NOM recall for these models drops from 100\% to around 98\% and then becomes stable. In other words, the mild improvement in performance seen from architectures \textcolor{black}{A and C} in terms of validation accuracy is partly contributed by the improvement of GRG recall but at the expense of NOM recall. 

\begin{figure*}
    \centering
    \includegraphics[width=\textwidth]{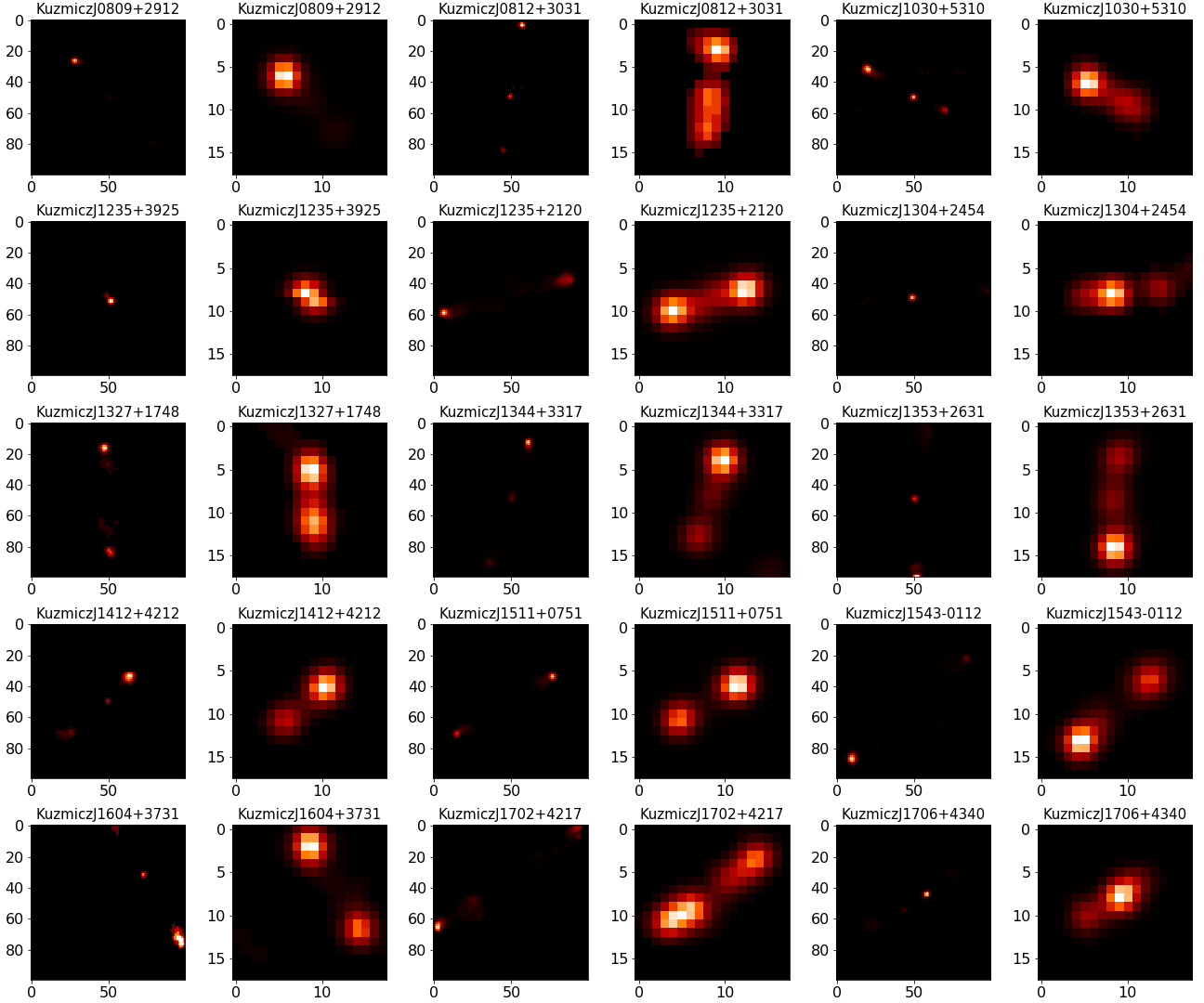}
    \caption{Example \textbf{GRG} images extracted from the \citet{kuzmicz2018} catalogue. Under the same object ID, the left image refers to its pre-processed FIRST image, while the right image is the pre-processed NVSS image of the object.}
    \label{fig:kuzmicz_test_sample}
\end{figure*}

\textcolor{black}{We also note that the architectures that use only NVSS images as their input tend to exhibit more stable training and result in a higher rate of correctly classified GRG samples. This can be seen in both Figure~\ref{fig:learning_curve_illustration_GRGNOM_A} and Table~\ref{tab:GRGNOM_A_model_performance_summary_no_dabhade}. Compared to Architectures B and D, which have only FIRST data as an image input, models trained with only NVSS data as image inputs (Architectures A and C) have higher GRG recall by 18 to 27 \% on average. This is likely to be caused by the GRG sample selection of GRGNOM-A, as shown in Figure~\ref{fig:kuzmicz_test_sample}, where GRG objects are cross-validated using NVSS images, and thus their radio components are more clearly visible compared to those of their FIRST image counterparts.}

\textcolor{black}{The inclusion of host galaxy redshift as an input feature boosts model performance regardless of architecture. For architectures with or without redshift (A \& C, B \& D, E \& F), it can be observed that the inclusion of redshift information causes a marginal improvement in model accuracy, AUC value and GRG class metrics, as seen in Table~\ref{tab:GRGNOM_A_model_performance_summary_no_dabhade}.}

\textcolor{black}{Although the selection of image inputs gives different performances when using single image input models, using both image inputs generally contribute to better classification results when they are imported together. Both with and without host galaxy redshift information or the presence of Inception modules, we found that architectures E, F and G outperform the single image input models across model AUC values and GRG class precision, along with similar or better model test accuracies as compared to those single image inputs. This is consistent with what has been found from other non-astronomical applications, that multi-branched approaches can boost model performance compared to the classical single input approach \citep[e.g.][]{2017arXiv171005477L,2020NatSR..10.9486G}. However, it is also noteworthy that the inclusion of multiple image inputs on average decreases GRG class recall by 7\%, implying that such multi-branch approaches should be treated with caution when the objective of the classifier is to identify as many GRGs as possible.}

\textcolor{black}{When we introduce the Inception module with dimension reduction to the model, we do not find a significant improvement in model performance when testing against the GRGNOM-A test sample. However, we find that this architecture performs differently when trained and test on GRGNOM-B, and this is discussed in more detail in the following section.}

\begin{table*}
    \begin{center}
    \begin{tabular}{c|c|c|c|c|c|c|c}
    \hline
    Architecture (IC regularization)     & \textbf{A} & \textbf{B} & \textbf{C} & \textbf{D} & \textbf{E} & \textbf{F} & \textbf{G} \\ 
    Input data  & NVSS &  FIRST  &  NVSS \& $z$  &  FIRST \& $z$  &  NVSS \& FIRST  & NVSS \& FIRST \& $z$ & NVSS \& FIRST \& $z$ \\\hline
    Accuracy (\%)  & 96.7 $\pm$ 0.5    & 95.7 $\pm$ 1.4    & 97.1 $\pm$ 0.6   & 96.4 $\pm$ 1.3    & 97.2 $\pm$ 0.9   & 97.4 $\pm$ 0.7   & 96.9 $\pm$ 0.9  \\
    AUC  & 0.948 $\pm$ 0.018   & 0.944 $\pm$ 0.027   & 0.953 $\pm$ 0.016   &  0.949 $\pm$ 0.027  & 0.974 $\pm$ 0.012    & 0.971 $\pm$ 0.014  & 0.969 $\pm$ 0.014 \\
    Train loss  & 0.080 $\pm$ 0.008    & 0.071 $\pm$ 0.014    & 0.067 $\pm$ 0.008   & 0.056 $\pm$ 0.012    & 0.070 $\pm$ 0.007   & 0.070 $\pm$ 0.008   & 0.074 $\pm$ 0.010  \\ 
    Test loss  & 0.107 $\pm$ 0.017    & 0.156 $\pm$ 0.022    & 0.105 $\pm$ 0.011   & 0.143 $\pm$ 0.032   & 0.103 $\pm$ 0.016   & 0.103 $\pm$ 0.027   & 0.100 $\pm$ 0.012  \\
    Generalization gap  & 0.027 $\pm$ 0.018    & 0.085 $\pm$ 0.026    & 0.038 $\pm$ 0.013   & 0.088 $\pm$ 0.034    & 0.033 $\pm$ 0.017   & 0.033 $\pm$ 0.028   & 0.026 $\pm$ 0.016  \\ \hline
    Precision (GRG) & 0.727 $\pm$ 0.045 & 0.738 $\pm$ 0.130   & 0.777 $\pm$ 0.056   & 0.793 $\pm$ 0.117    & 0.821 $\pm$ 0.071    & 0.848 $\pm$ 0.060    & 0.800 $\pm$ 0.080  \\
    Recall (GRG) & 0.796 $\pm$ 0.049   & 0.521 $\pm$ 0.170   & 0.783 $\pm$ 0.067   & 0.603 $\pm$ 0.143  & 0.723 $\pm$ 0.108   & 0.732 $\pm$ 0.089    & 0.698 $\pm$ 0.103  \\
    F1 score (GRG)  & 0.759 $\pm$ 0.036   & 0.603 $\pm$ 0.155  & 0.778 $\pm$ 0.048    & 0.679 $\pm$ 0.129   & 0.765 $\pm$ 0.080   & 0.783 $\pm$ 0.065   & 0.741 $\pm$ 0.080  \\ \hline
    
    Architecture (BN regularization)     & \textbf{A} & \textbf{B} & \textbf{C} & \textbf{D} & \textbf{E} & \textbf{F} & \textbf{G} \\ 
    Input data  & NVSS &  FIRST  &  NVSS \& $z$  &  FIRST \& $z$  &  NVSS \& FIRST  & NVSS \& FIRST \& $z$ & NVSS \& FIRST \& $z$ \\\hline
    Accuracy (\%)  & 96.9 $\pm$ 0.5    & 97.0 $\pm$ 0.8    & 97.2 $\pm$ 0.5   & 96.7 $\pm$ 0.9    & 97.7 $\pm$ 0.6   & 97.9 $\pm$ 0.6   & 97.7 $\pm$ 0.6  \\
    AUC  & 0.957 $\pm$ 0.015   & 0.955 $\pm$ 0.025   & 0.967 $\pm$ 0.011   &  0.953 $\pm$ 0.021  & 0.987 $\pm$ 0.008    & 0.985 $\pm$ 0.010  & 0.989 $\pm$ 0.005 \\
    Train loss  & 0.057 $\pm$ 0.004    & 0.026 $\pm$ 0.011    & 0.040 $\pm$ 0.005   & 0.027 $\pm$ 0.012    & 0.033 $\pm$ 0.005   & 0.031 $\pm$ 0.005   & 0.0253 $\pm$ 0.006  \\ 
    Test loss  & 0.117 $\pm$ 0.010    & 0.149 $\pm$ 0.042    & 0.107 $\pm$ 0.012   & 0.172 $\pm$ 0.033   & 0.080 $\pm$ 0.014   & 0.073 $\pm$ 0.016   & 0.088 $\pm$ 0.022  \\
    Generalization gap  & 0.061 $\pm$ 0.010    & 0.123 $\pm$ 0.044    & 0.067 $\pm$ 0.013   & 0.146 $\pm$ 0.035    & 0.047 $\pm$ 0.015   & 0.042 $\pm$ 0.016   & 0.063 $\pm$ 0.022  \\ \hline
    Precision (GRG) & 0.748 $\pm$ 0.042 & 0.837 $\pm$ 0.070   & 0.775 $\pm$ 0.049   & 0.812 $\pm$ 0.082    & 0.843 $\pm$ 0.056    & 0.858 $\pm$ 0.054    & 0.839 $\pm$ 0.059  \\
    Recall (GRG) & 0.789 $\pm$ 0.066   & 0.670 $\pm$ 0.113   & 0.801 $\pm$ 0.053   & 0.641 $\pm$ 0.103  & 0.791 $\pm$ 0.079   & 0.808 $\pm$ 0.071    & 0.798 $\pm$ 0.087  \\
    F1 score (GRG)  & 0.766 $\pm$ 0.042   & 0.740 $\pm$ 0.088  & 0.786 $\pm$ 0.036    & 0.712 $\pm$ 0.085   & 0.814 $\pm$ 0.055   & 0.830 $\pm$ 0.051   & 0.814 $\pm$ 0.055  \\ \hline 
    
    \end{tabular}
    \end{center}
    \caption{Summary of model performance metrics for all architectures trained and tested with the GRGNOM-A dataset. Each model architecture listed are trained 10 times on the GRGNOM-A training sets using independent Xavier initializations, and perform 10 independent model evaluations on each trained model.}
    \label{tab:GRGNOM_A_model_performance_summary_no_dabhade}
\end{table*}

\subsubsection{Models trained with GRGNOM-B}
\label{sec:model_trained_with_GRGNOM_B}

\begin{figure*}
\setlength{\unitlength}{1cm}
\includegraphics[width=\textwidth]{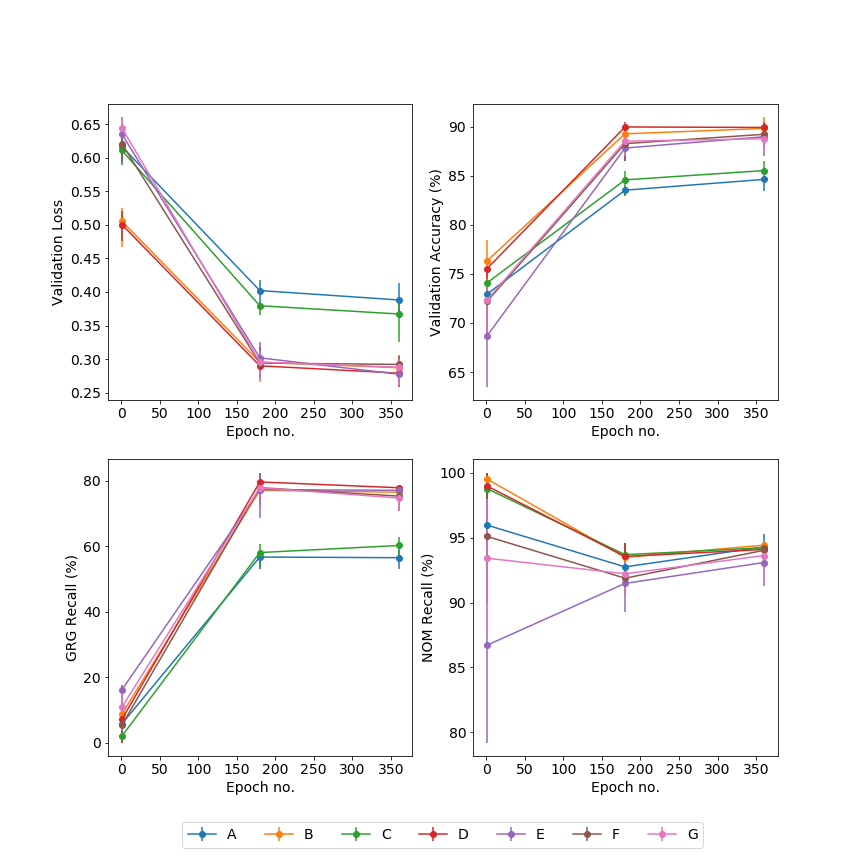}
\caption{The averaged learning curve of the model architectures trained and validated with GRGNOM-B. Assuming normal distribution, the asymmetric errors of each data point on the diagram has covered 60\% of data distribution.}
\label{fig:learning_curve_illustration_GRGNOM_B}
\end{figure*}

By comparing Table~\ref{tab:GRGNOM_A_model_performance_summary_no_dabhade} and Table~\ref{tab:GRGNOM_B_model_performance_summary_no_kuzmicz}, it can be seen that models trained using the GRGNOM-B data set are able to make more stable predictions. The inclusion of the \citet{dabhadelotss} data in the training set lowers the class imbalance ratio from 14:1 in GRGNOM-A to around 3:1 in this data set. With more GRG examples in the training set, models are able to learn more quickly and make more stable predictions, see  Figure~\ref{fig:learning_curve_illustration_GRGNOM_B}. 

The biggest difference between Figure~\ref{fig:learning_curve_illustration_GRGNOM_A} and Figure~\ref{fig:learning_curve_illustration_GRGNOM_B} is the reversal of model performance differences between single image input models trained with NVSS images and FIRST images. Compared with Architectures A and C, Architectures B and D have lower test losses and higher test accuracies after 180 epochs of training. The largest contribution to this difference can be attributed to data sample selection. The 101 \citet{dabhadelotss} samples in the GRGNOM-B training set are identified from LoTSS survey maps with an angular resolution of 6$''$, and consequently the source morphology of these objects is found to be much closer to that of the FIRST survey with an angular resolution of 5.4$''$, rather than the lower resolution NVSS survey. Such similarity in angular resolution will contribute to model performance: once FIRST image inputs are imported, models trained with GRGNOM-B data samples receive F1 scores higher than \textcolor{black}{0.76} on average, see Table~\ref{tab:GRGNOM_B_model_performance_summary_no_kuzmicz}. Moreover, models trained using only FIRST images as inputs have GRG Recall/Precision values \textcolor{black}{$\ge$19.7/6.8\%} higher than those trained with equivalent NVSS images (A \& B).

\textcolor{black}{The inclusion of host galaxy redshift results in similar (B \& D) or improved (A \& C, E \& F) model performance when trained and tested using the GRGNOM-B data set. The influence of the multi-branch network approach, however, appears to behave differently. Compared to Architectures A and B, Architecture E was found to have similar or poorer model performance.}

Interestingly, the inclusion of the Inception modules also seems to mildly improve model performance when testing with the GRGNOM-B data set. This is perhaps also due to the higher resolution sample selection for this data set. The extra network parameters are able to learn more complex source morphology features from these samples, and thus have slightly boosted model performance relative to other architectures. 

\begin{figure*}
\setlength{\unitlength}{1cm}
\includegraphics[width=\textwidth]{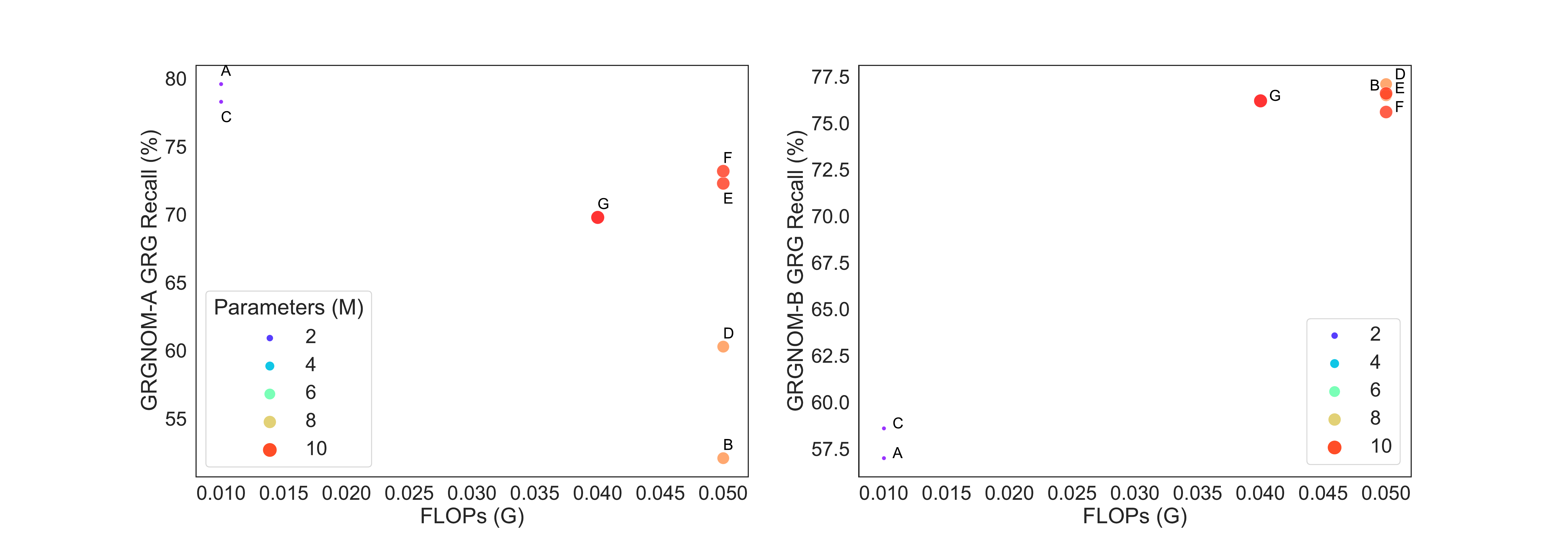}
\caption{Left: Model computational complexity vs. GRG Class Recall \textcolor{black}{for an architecture regularized via IC, trained and tested on the GRGNOM-A dataset.} Model computational complexity is evaluated by computing the Giga floating point operations per second (G-FLOPs) shown on the x-axis. Model GRG Recall is shown on the y-axis as a percentage (\%) \textcolor{black}{given that the metric is of most importance if one aims to generate a comprehensive GRG catalogue from the GRGNOM-A testset using the models in our work}. The size of the circles on the diagram indicates the number of trainable model parameters for each architecture (A to G) considered in this work in units of $10^6$. Right: The same diagram for each architecture trained and tested on the GRGNOM-B dataset.}
\label{fig:flops_vs_recalls}
\end{figure*}

\textcolor{black}{Other than model performance evaluation, we also measure the model computational complexity of each architecture. This is achieved by measuring the number of floating point operations for a single instance of a forward pass through a given model \citep{2021MNRAS.503.1828B}. It can be seen from Figure~\ref{fig:flops_vs_recalls} that the Architecture G in our work has reduced model complexity by 0.01 Giga-FLOating Point operations (G-FLOPs) for models trained on both GRGNOM-A and GRGNOM-B compared to Architecture F.}

\begin{table*}
    \begin{center}
    \begin{tabular}{c|c|c|c|c|c|c|c}
    \hline
    Architecture (IC regularization)     & \textbf{A} & \textbf{B} & \textbf{C} & \textbf{D} & \textbf{E} & \textbf{F} & \textbf{G} \\ 
    Input data  & NVSS &  FIRST  &  NVSS \& $z$  &  FIRST \& $z$  &  NVSS \& FIRST  & NVSS \& FIRST \& $z$ & NVSS \& FIRST \& $z$ \\\hline
    Accuracy (\%) & 84.6 $\pm$ 1.0   & 89.9 $\pm$ 1.2   & 85.1 $\pm$ 1.2 & 89.9 $\pm$ 1.3  & 88.9 $\pm$ 1.3  & 89.1 $\pm$ 1.4  & 89.3 $\pm$ 1.3 \\
    AUC  & 0.873 $\pm$ 0.019 & 0.928 $\pm$ 0.011  & 0.891 $\pm$ 0.017  & 0.925 $\pm$ 0.011  & 0.926 $\pm$ 0.010   & 0.928 $\pm$ 0.010 & 0.931 $\pm$ 0.011 \\
    Train loss  & 0.285 $\pm$ 0.008    & 0.212 $\pm$ 0.015    & 0.254 $\pm$ 0.006   & 0.205 $\pm$ 0.010    & 0.223 $\pm$ 0.008   & 0.218 $\pm$ 0.013   & 0.222 $\pm$ 0.012  \\ 
    Test loss  & 0.388 $\pm$ 0.028    & 0.287 $\pm$ 0.016    & 0.367 $\pm$ 0.028   & 0.279 $\pm$ 0.016    & 0.278 $\pm$ 0.015   & 0.292 $\pm$ 0.017   & 0.288 $\pm$ 0.021  \\ 
    Generalization gap  & 0.104 $\pm$ 0.029    & 0.075 $\pm$ 0.022    & 0.113 $\pm$ 0.029   & 0.0741 $\pm$ 0.019    & 0.055 $\pm$ 0.017   & 0.075 $\pm$ 0.022   & 0.066 $\pm$ 0.024  \\ \hline
    Precision (GRG)  & 0.766 $\pm$ 0.033  & 0.826 $\pm$ 0.037  & 0.778 $\pm$ 0.034   & 0.823 $\pm$ 0.035   & 0.792 $\pm$ 0.036   & 0.806 $\pm$ 0.038 & 0.809 $\pm$ 0.037 \\
    Recall (GRG) & 0.570 $\pm$ 0.027  & 0.765 $\pm$ 0.035   & 0.586 $\pm$ 0.037   & 0.771 $\pm$ 0.033  & 0.766 $\pm$ 0.033  & 0.756 $\pm$ 0.033  & 0.762 $\pm$ 0.030 \\
    F1 score (GRG) & 0.653 $\pm$ 0.023  & 0.794 $\pm$ 0.024  & 0.668 $\pm$ 0.030  & 0.796 $\pm$ 0.026  & 0.778 $\pm$ 0.026   & 0.780 $\pm$ 0.027  & 0.784 $\pm$ 0.025  \\ \hline
    
    Architecture (BN regularization)     & \textbf{A} & \textbf{B} & \textbf{C} & \textbf{D} & \textbf{E} & \textbf{F} & \textbf{G} \\ 
    Input data  & NVSS &  FIRST  &  NVSS \& $z$  &  FIRST \& $z$  &  NVSS \& FIRST  & NVSS \& FIRST \& $z$ & NVSS \& FIRST \& $z$ \\\hline
    Accuracy (\%) & 86.2 $\pm$ 1.2   & 90.2 $\pm$ 1.5   & 86.5 $\pm$ 1.0 & 90.5 $\pm$ 1.9  & 91.1 $\pm$ 1.0  & 90.7 $\pm$ 1.1  & 91.4 $\pm$ 1.2 \\
    AUC  & 0.9 $\pm$ 0.011 & 0.935 $\pm$ 0.010  & 0.915 $\pm$ 0.010  & 0.934 $\pm$ 0.011  & 0.941 $\pm$ 0.010   & 0.941 $\pm$ 0.009 & 0.942 $\pm$ 0.009 \\
    Train loss  & 0.229 $\pm$ 0.008    & 0.156 $\pm$ 0.008    & 0.197 $\pm$ 0.007   & 0.158 $\pm$ 0.012    & 0.139 $\pm$ 0.011   & 0.140 $\pm$ 0.013   & 0.126 $\pm$ 0.016  \\ 
    Test loss  & 0.346 $\pm$ 0.014    & 0.286 $\pm$ 0.023    & 0.330 $\pm$ 0.017   & 0.280 $\pm$ 0.026    & 0.265 $\pm$ 0.020   & 0.275 $\pm$ 0.020   & 0.283 $\pm$ 0.019  \\ 
    Generalization gap  & 0.118 $\pm$ 0.017    & 0.131 $\pm$ 0.024    & 0.133 $\pm$ 0.019   & 0.122 $\pm$ 0.028    & 0.126 $\pm$ 0.022   & 0.136 $\pm$ 0.024   & 0.157 $\pm$ 0.025  \\ \hline
    Precision (GRG)  & 0.812 $\pm$ 0.030  & 0.825 $\pm$ 0.061  & 0.818 $\pm$ 0.024   & 0.836 $\pm$ 0.065   & 0.862 $\pm$ 0.026   & 0.863 $\pm$ 0.043 & 0.874 $\pm$ 0.033 \\
    Recall (GRG) & 0.597 $\pm$ 0.032  & 0.794 $\pm$ 0.054   & 0.604 $\pm$ 0.042   & 0.787 $\pm$ 0.034  & 0.774 $\pm$ 0.038  & 0.761 $\pm$ 0.042  & 0.777 $\pm$ 0.035 \\
    F1 score (GRG) & 0.688 $\pm$ 0.029  & 0.806 $\pm$ 0.027  & 0.694 $\pm$ 0.028  & 0.809 $\pm$ 0.031  & 0.815 $\pm$ 0.023   & 0.807 $\pm$ 0.024  & 0.822 $\pm$ 0.025  \\ \hline

    \end{tabular}
    \end{center}
    \caption{Summary of model performance metrics for all architectures trained and tested with the GRGNOM-B dataset. Each model architecture listed are trained 10 times on the GRGNOM-B training sets using independent Xavier initializations, and perform 10 independent model evaluations on each trained model.}
    \label{tab:GRGNOM_B_model_performance_summary_no_kuzmicz}
\end{table*}

\subsubsection{Dataset Shift}
\label{sec:dataset_shift}

\textcolor{black}{To this point we have evaluated the models in this work using test data taken from the same underlying data set as the training data, either GRGNOM-A or GRGNOM-B.} \textcolor{black}{However, it is also important to evaluate the generalization ability of a model when there exist small differences between underlying data distributions, a phenomenon referred to as dataset shift.}


\textcolor{black}{A model with good generalization ability is able to perform well on new inputs unseen during the model training phase \citep{GoodBengCour16}. However, the use of this term should be treated with caution when the dataset foundations do not follow i.i.d assumptions: i.e. that samples in the training and test sets are independent from each other, and share an identical underlying distribution \citep{GoodBengCour16}. One would then need to clarify that the model's generalization ability is evaluated under dataset shift.}

\textcolor{black}{Almost all quantitative metrics used to evaluate a model's generalization ability are usually based on an assumption that the training set and the testing set share an identical joint generating distribution of inputs and outputs \citep{10.5555/1462129}. However, it is easy for this assumption to be violated. For instance, such an assumption will not hold when only the distribution of inputs is changed between the training and test set, a situation known as simple covariate shift, or alternatively when only the data class distribution changes (prior probability shift), or the training data samples do not accurately represent the full data distribution of the test samples (sample selection bias). A difference in the data source between the training and test set could also violate this assumption (source component shift). These scenarios can all be summarized as examples of dataset shift \citep{10.5555/1462129}, which is almost inevitable when applying a model to make predictions on unseen data in a real world scenario.}

\textcolor{black}{In order to evaluate model generalization ability when dataset shift exists,} we now consider models trained using GRGNOM-A and tested using data from the 51 GRG samples in the GRGNOM-B test set. We also test these models with another 149 RGZ DR1 samples of class NOM that are not found in either training set. We refer to the resulting test set as GRGNOM-Gen. \textcolor{black}{When comparing the GRGNOM-Gen test set with the GRGNOM-A train set, it is clear that both sample selection bias and source component shift are likely to be present, and so we expect to see these examples of dataset shift manifest in differing model performance.} We do not consider the opposite approach, as the GRG samples in the GRGNOM-A test set have been included in the GRGNOM-B training set. The evaluation metrics from these tests can be seen in Table~\ref{tab:GRGNOM_A_model_performance_summary_no_kuzmicz}.

\textcolor{black}{From Table~\ref{tab:GRGNOM_A_model_performance_summary_no_kuzmicz} it can be seen that the previously observed advantage of including multi-domain data does not hold when testing on GRGNOM-Gen, resulting in a decrease of 2.2\% in model test accuracy, 0.029 in model AUC value and 9.5\% decrease in GRG recall (e.g. D \& E).} More generally, it can be seen that when making predictions on these test samples, models trained with GRGNOM-A perform comparatively less well in terms of general model metrics than those directly trained on the GRGNOM-B data set. A similar situation also occurs when looking at GRG recall and $\rm F_{1}$ score: \textcolor{black}{even the best performing Architecture~D can only provide a GRG recall of 0.419 $\pm$ 0.043. On the other hand, the same architecture has a GRG precision of 88.7\% on average,} implying that the model is able to identify NOM class objects well. In other words, although these models achieve higher GRG classification precision when compared to those trained using the GRGNOM-B dataset, they are unable to reach a comparably high classification completeness. In order to find the majority of the GRGs in the \citet{dabhadelotss} sample, it is essential to have some of the \citet{dabhadelotss} samples included in the model training set. 

\begin{table*}
    \begin{center}
    \begin{tabular}{c|c|c|c|c|c|c|c}
    \hline
    Architecture (IC regularization)    & \textbf{A} & \textbf{B} & \textbf{C} & \textbf{D} & \textbf{E} & \textbf{F} & \textbf{G} \\ 
    Input data  & NVSS &  FIRST  &  NVSS \& $z$  &  FIRST \& $z$  &  NVSS \& FIRST  & NVSS \& FIRST \& $z$ & NVSS \& FIRST \& $z$ \\\hline
    Accuracy (\%) & 80.9 $\pm$ 0.8  & 83.6 $\pm$ 1.3   & 81.3 $\pm$ 0.9    & 83.8 $\pm$ 1.3   & 81.6 $\pm$ 1.1  & 81.7 $\pm$ 1.0   & 81.4 $\pm$ 0.9 \\
    AUC  & 0.787 $\pm$ 0.021   & 0.828 $\pm$ 0.018   & 0.805 $\pm$ 0.016   & 0.833 $\pm$ 0.020   & 0.804 $\pm$ 0.023  & 0.805 $\pm$ 0.023   & 0.808 $\pm$ 0.021 \\\hline
    Precision (GRG)& 0.848 $\pm$ 0.047  & 0.883 $\pm$ 0.047  & 0.882 $\pm$ 0.044  & 0.887 $\pm$ 0.044   & 0.878 $\pm$ 0.050    & 0.884 $\pm$ 0.051   & 0.866 $\pm$ 0.046 \\
    Recall (GRG) & 0.305 $\pm$ 0.026  & 0.415 $\pm$ 0.047  & 0.308 $\pm$ 0.031   & 0.419 $\pm$ 0.043  & 0.324 $\pm$ 0.038   & 0.327 $\pm$ 0.031  & 0.320 $\pm$ 0.033 \\
    F1 score (GRG) & 0.448 $\pm$ 0.031   & 0.562 $\pm$ 0.045 &  0.456 $\pm$ 0.037 & 0.568 $\pm$ 0.044 & 0.472 $\pm$ 0.042  & 0.477 $\pm$ 0.036  & 0.466 $\pm$ 0.036 \\ \hline

    Architecture (BN regularization)    & \textbf{A} & \textbf{B} & \textbf{C} & \textbf{D} & \textbf{E} & \textbf{F} & \textbf{G} \\ 
    Input data  & NVSS &  FIRST  &  NVSS \& $z$  &  FIRST \& $z$  &  NVSS \& FIRST  & NVSS \& FIRST \& $z$ & NVSS \& FIRST \& $z$ \\\hline
    Accuracy (\%) & 81.3 $\pm$ 0.6  & 84.1 $\pm$ 1.1   & 81.4 $\pm$ 0.6    & 84.1 $\pm$ 1.2   & 82.7 $\pm$ 0.8  & 82.5 $\pm$ 0.7   & 82.6 $\pm$ 0.8 \\
    AUC  & 0.767 $\pm$ 0.027   & 0.841 $\pm$ 0.028   & 0.774 $\pm$ 0.019   & 0.834 $\pm$ 0.037   & 0.808 $\pm$ 0.021  & 0.812 $\pm$ 0.023   & 0.823 $\pm$ 0.016 \\\hline
    Precision (GRG)& 0.858 $\pm$ 0.032  & 0.869 $\pm$ 0.048  & 0.866 $\pm$ 0.027  & 0.873 $\pm$ 0.044   & 0.884 $\pm$ 0.028    & 0.885 $\pm$ 0.028   & 0.882 $\pm$ 0.026 \\
    Recall (GRG) & 0.320 $\pm$ 0.022  & 0.446 $\pm$ 0.045  & 0.318 $\pm$ 0.023   & 0.441 $\pm$ 0.047  & 0.370 $\pm$ 0.027   & 0.360 $\pm$ 0.029  & 0.365 $\pm$ 0.034 \\
    F1 score (GRG) & 0.465 $\pm$ 0.024   & 0.587 $\pm$ 0.039 &  0.465 $\pm$ 0.025 & 0.584 $\pm$ 0.042 & 0.521 $\pm$ 0.029  & 0.511 $\pm$ 0.030  & 0.515 $\pm$ 0.035 \\ \hline
    
    \end{tabular}
    \end{center}
    \caption{Summary of model performance metrics for all architectures trained with the GRGNOM-A dataset and tested with the model generalization test set described in Section~\ref{sec:dataset_shift}. Each model architecture listed are trained 10 times on the GRGNOM-A training sets using independent Xavier initializations, and perform 10 independent model evaluations on each trained model.}
    \label{tab:GRGNOM_A_model_performance_summary_no_kuzmicz}
\end{table*}

\subsubsection{Angular Size Distance vs. Host galaxy redshift}
\label{sec:d_A}

A consideration when introducing host galaxy redshift as an input feature is that the relationship between host galaxy redshift and angular size distance, $D_{A}$, is not linear. As an experiment, we used the equivalent $D_{A}$ in Gpc to replace host galaxy redshift when training Architecture~F using the GRGNOM-A data set. 
\textcolor{black}{The resulting models have an average AUC of 0.970 $\pm$ 0.016, slightly lower than, but not significantly different from, that found when using $z$ directly. When looking at GRG classification performance, the $D_A$ alternative returns a GRG Precision of 0.814$\pm$0.069 and a GRG Recall of 0.707$\pm$0.092. Comparing these metrics with those in Table~\ref{tab:GRGNOM_A_model_performance_summary_no_dabhade}, which use $z$ directly, it can be seen that the results are slightly poorer. This suggests that the network architecture already has sufficient capacity in its trainable parameters to learn the $z-D_{A}$ relationship, or an approximation of it.}

\subsection{BN vs. IC}

\textcolor{black}{In order to compare the utility of the IC and BN layers, we performed a comprehensive model performance evaluation on all architectures that used either IC or BN for regularization, see Table \ref{tab:GRGNOM_A_model_performance_summary_no_dabhade} - \ref{tab:GRGNOM_A_model_performance_summary_no_kuzmicz}). In general, both approaches are able to prevent our models from over-fitting. The benefit of using BN for regularization is that it boosted model test accuracy, AUC, GRG recall in most cases by 1 to 3\% compared with those models that used IC. On the other hand, models using IC share smaller generalization gaps, i.e. the gap between the training and test loss values. The decrease in training loss for models regularized by BN is the major factor causing this difference, while these models typically have comparable test loss to those that used IC as a regularization method.}

\textcolor{black}{When we dive into specific architectures or datasets, we find several model performance differences between the two approaches. When we looked at those models which adopted BN regularization, trained and tested on GRGNOM-B, we found that they experienced a performance advantage (both for test accuracy and AUC value) from having multiple image inputs. It is true though that the model GRG recall behaves more poorly than for those trained with single image inputs (B/D and E), regardless of regularization method.}

\textcolor{black}{Moreover, in our previous discussion of different architectures regularized by IC, we noted that including redshift information provided similar or improved model performance for those trained on GRGNOM-A and tested on the GRGNOM-Gen dataset, depending on architecture. However, we did not observe such a  comprehensive improvement when we tested the models using BN for regularization on GRGNOM-Gen: models trained with redshift gained a 0.1-0.8\% improvement in GRG Precision, at the expense of a 0.2-1.0\% decrease in GRG Recall. Finally, when comparing Architectures F and G, we note that the inclusion of inception modules leads to slightly better performance (0.5\% improvement in GRG recall) when testing on the GRGNOM-Gen dataset. Consequently, we conclude that IC is a non-harmful regularization method, but does not show significant benefit in this specific application. Users would need to compare its regularization ability with other methods when deciding model regularization strategies.}

\subsection{Common features shared by the misidentified samples}
\label{sec:common_features_shared_by_mis_iden_sam}

The model evaluation we have presented so far is based on a simple assumption: that the GRGNOM-A/B data sets are fully understood, reliable and confidently labelled. Yet it is unclear whether frequently misclassified objects in our models share common features. \textcolor{black}{In this section and in Section~\ref{sec:further_explanation}, we consider the models using IC as a regularization method and trained with the GRGNOM-B dataset as examples. By applying each of these models to all samples in the \textcolor{black}{GRGNOM-B test set}, we are able to identify the \textcolor{black}{GRGNOM-B} test samples that have a misclassification rate of $\ge$50\% for all of the 7 architectures used in this work.} These samples are summarised in Table~\ref{tab:mis_summary_new}. In general,  our models mistakenly identify GRGs as class NOM more frequently than the reverse. This is unsurprising since both the GRGNOM-A and GRGNOM-B training samples contain a much higher number of NOM-type objects.

A potential data `trap' in the GRGNOM-B data sets comes from the \citet{dabhadelotss} samples. These objects were identified using the 151\,MHz LoTSS survey \citep{2019A&A...622A...1S}. Considering that $S \propto \nu^{\alpha}$, where $\rm \alpha = -0.7$ for optically thin synchrotron emission, sources will be brighter at 151\,MHz compared to their NVSS or FIRST counterparts at 1.4\,GHz. In addition, the median rms of LoTSS is 71\,$\mu$Jy\,beam$^{-1}$, lower than half that of the FIRST survey and around 16\% of the NVSS sensitivity of 0.45\,mJy\,beam$^{-1}$. This means that LoTSS will be more sensitive to faint radio emission, and that some radio structures present in LoTSS images might be missed or resolved out in the equivalent NVSS and/or FIRST images. Besides the `trap', the aforementioned image specifications, pre-processing choices, selection of input domains (NVSS images, FIRST images, host galaxy redshifts), and selection of architectures could also result in differences in model performance.

In order to investigate these frequent misclassifications we use the data traceability built into our data set, see Section~\ref{sec:data_sample}. Similarly traceable data sets have been built and implemented for a number of recent deep learning studies \citep[e.g.][]{2019MNRAS.482.1211W,2020MNRAS.491.1554W} and their data traceability used to explain why some samples are mistakenly identified in a frequent manner \citep[e.g.][]{2019MNRAS.482.1211W}. In this case, the sources listed in  Table.~\ref{tab:mis_summary_new} were traced using their unique object IDs, see Section~\ref{sec:data_formating}. We then analysed these cases of object  misclassification by looking at their NVSS and FIRST pre-processed images as well as the host galaxy redshift in each case.
\begin{table}
    \centering
    \begin{center}
    \begin{tabular}{c|c}
    \hline
    Object ID     & $\ge$50\% Mistakenly Identified Architectures   \\ \hline
    Dabhade230	 &  All \\
    Dabhade217	 &  All \\
    Dabhade198	 &  All \\
    Dabhade173	 &  All \\
    Dabhade237	 &  All \\
    Dabhade186	 &  All \\ 
    Dabhade216	 &  All \\
    Dabhade193	 &  All \\ \hline
    RGZJ080417.6+320250    &	A,B,C,E,F,G \\ \hline
    RGZJ075855.6+360246	   &    A,B,C,D,E,F \\ \hline
    Dabhade204             &	A,C,E,F,G \\
    RGZJ075030.7+525022	   &    A,C,E,F,G \\
    RGZJ080448.0+081254	   &    A,C,E,F,G \\
    RGZJ075539.6+160158	   &    A,C,E,F,G \\ \hline
    RGZJ075157.7+212049	   &    B,D,E,F,G \\ \hline
    RGZJ080427.8+132930	   &    A,D,E,F   \\ \hline
    RGZJ075812.7+190043	   &    B,D,E,G   \\ \hline
    Dabhade185	           &    A,B,C,D   \\ \hline
    Dabhade221	           &    B,D,F     \\ \hline
    Dabhade201	           &    A,B,D     \\ \hline
    Dabhade197	           &    A,C,G     \\ \hline
    Dabhade214	           &    B,F,G     \\ \hline
    Dabhade199	           &    A,C       \\
    Dabhade227	           &    A,C       \\
    Dabhade206	           &    A,C       \\
    Dabhade226	           &    A,C       \\
    RGZJ074627.1+174337	   &    A,C       \\
    RGZJ074720.7+335008	   &    A,C       \\
    Dabhade229	           &    A,C       \\
    Dabhade231	           &    A,C       \\
    Dabhade220	           &    A,C       \\
    Dabhade209	           &    A,C       \\
    Dabhade163	           &    A,C       \\ \hline
    RGZJ080404.5+153334    &	B,D       \\
    RGZJ075306.1+121504	   &    B,D       \\ \hline
    Dabhade210	           &    E,G       \\ \hline
    RGZJ080402.5+452258    &	E         \\
    RGZJ075620.0+301630	   &    E         \\ \hline
    \end{tabular}
    \end{center}
    \caption{A summary of frequent mistakenly identified GRGNOM-B testing samples in this work. A sample would be included in this table if it has over 50\% rate to be mistakenly identified in at least one architecture adopted in this work.}
    \label{tab:mis_summary_new}
\end{table}

\begin{figure*}
    \centering
    \includegraphics[width=0.85\textwidth]{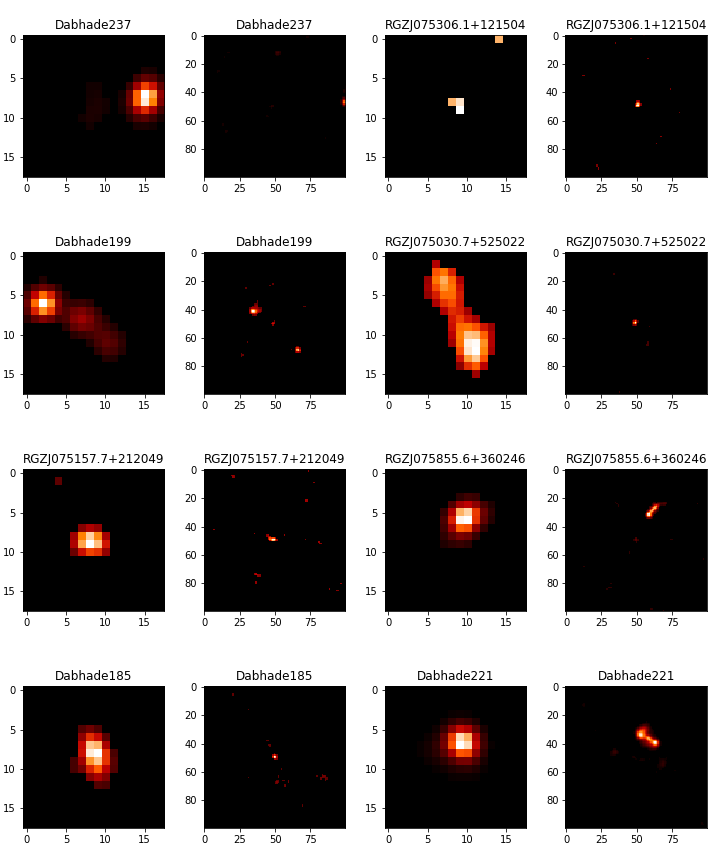}
    \caption{A summary of misidentified GRGNOM-B testing samples, which represent the typical types of misclassification described in Section~\ref{sec:common_features_shared_by_mis_iden_sam}. Under the same object ID, the left image refers to its pre-processed NVSS image, while the right image is the pre-processed FIRST image of the object.}
    \label{fig:mistaken_identified_sample_example}
\end{figure*}

\subsubsection{Low surface brightness}

As described in Section~\ref{sec:data_sample}, the sigma-clipping performed during image pre-processing will replace all pixels with values lower than the local 3-$\sigma_{\rm rms}$ level with zeros, resulting in faint objects that have only a  mild luminosity difference with the noise background becoming even fainter in the normalized image when a secondary source is present in the field of view. For example, the source \textcolor{black}{Dabhade\,237} was faint but visible at 151\,MHz in the original LOFAR data \citep[FigureA.8;][]{dabhadelotss}. However, the both the radio core and the nearby radio lobes seem to be too faint to be clearly identified in the pre-processed FIRST and NVSS maps used in this work. The models are therefore unable to find a GRG like object in the image, but instead identify a secondary source at the edge of the field to be class NOM. This suggests that using data samples selected at different frequencies for model training/testing should be treated with caution.

\subsubsection{Input Domain Selection}

\textcolor{black}{In general, multi-domain approaches are expected to be able to help a network to cross-validate its prediction results. In this work, 39\% of the frequently misidentified objects listed in Table~\ref{tab:mis_summary_new} have their identification corrected when both image domains are imported. The sources RGZ\,J075306.1+121504 and Dabhade\,199 are two representative examples: RGZ\,J075306.1+121504 is frequently misidentified without the inclusion of its NVSS image. In spite of the slightly extended structure along with scattered faint background emissions in its FIRST image, its NVSS image retains clear compact structure, helping the networks to correctly identify the object class. On the other hand, Dabhade\,199, whilst still having segments of its lobes resolved out in the FIRST image, has a clearly extended central source. In this case, the lower resolution NVSS image leads to confusion between two objects: Dabhade\,199 as a GRG and another compact source, which makes it difficult for the network to identify Dabhade\,199 as a GRG with confidence; however, with the help of the additional information from the FIRST image, the networks are able to classify this source correctly in most cases. }

\textcolor{black}{Conversely, the use of multiple image domains may also have a negative impact on model performance. Two examples of this are RGZ\,J075030.7+525022 and RGZ\,J075157.7+212049. While RGZ\,J075030.7+525022 looks compact in its FIRST image, the object shows a well-extended morphology in its NVSS map. Such differences in sample morphology lead to architectures that include the NVSS image input being unable to make a correct classification. On the other hand, the source RGZ\,J075157.7+212049 shows an extended morphology along with significant scattered emission in its FIRST map, while its NVSS map is dominated by a single compact object. Architectures including FIRST image data as an input make incorrect identifications in this case.} \textcolor{black}{Considering these examples, it appears that sources with different numbers of radio components in different surveys are more likely to be misclassified.}

\subsubsection{Architecture bias}

\textcolor{black}{As well as the selection of input data, differences in network architecture will also contribute to model performance. For examples, RGZ\,J075855.6+360246 ($z=0.336$; see Figure~\ref{fig:mistaken_identified_sample_example}) was correctly identified only when the Inception module was introduced. Both of the NVSS and FIRST images for this source include a nearby well-extended object, while the target itself only appears in its FIRST image, where it appears compact and faint. Considering the uniform kernel size of 5 for the convolutional layers in the models of Architectures A to F, these models may have found it difficult to capture the morphological features of the small and faint object. However, by introducing an Inception module to a model with Architecture G, the network was able to capture features down to a 1-pixel scale. This perhaps explain why the object is correctly identified only when we use Architecture G.} \textcolor{black}{Given that 78.9\% of the misclassified objects in Table~\ref{tab:mis_summary_new} have at least one architecture producing a different result from the others, it appears that those sources where architectures disagree with each other are more likely to be misclassified.}

\subsection{Cases requiring further explanation}
\label{sec:further_explanation}

\textcolor{black}{Apart from the aforementioned examples, there are a number of frequently misclassified objects in Table~\ref{tab:mis_summary_new} that we found difficult to explain by simply looking at object sample data. For instance, Dabhade\,185 can be seen as a unique example of multi-domain cross validation. Its radio lobes in the FIRST image have been resolved out, but are unresolved in the NVSS image map. This object was classified as GRG when both NVSS or FIRST are used, which we find difficult to interpret.}

\textcolor{black}{Another issue we encounter when trying to explain model behavior occurs when there is more than one well-extended object in an image map. From Table~\ref{tab:mis_summary_new} it can be seen that the source Dabhade\,221 was consistently misclassified by three of the models in this work. In this case, via visual inspection we found that apart from the GRG, there is another well-extended radio source visible in the FIRST image, see Figure~\ref{fig:mistaken_identified_sample_example}. The translational equivariance of the convolutional layers does not require a source to be located at the centre of the image in order to be correctly classified and this is an example of such a circumstance. Which source emission finally contributes to the prediction remains uncertain.}

Analysis of such complex cases could perhaps benefit from the use of \textcolor{black}{eXplainable Artificial Intelligence (XAI)} tools \citep[e.g., SHAP; ][]{2017arXiv170507874L}. Rather than visualizing feature maps for each specific layer, these tools allow users to directly visualize which features the network as a whole has recognized from each sample image, and in some cases can evaluate their contributions to target class identification. Another possibility might be to use attention-gating, which produces attention maps that similarly  facilitate the interpretation of a classification choice as made by the model \citep[see e.g.][]{2021MNRAS.501.4579B}. 

\subsection{Comparison to other automated search methods}
\label{sec:proctor}

As described in Section~\ref{sec:automated}, a previous approach to automated GRG identification was made by \cite{2016ApJS..224...18P} using a decision tree based machine learning approach. Using source pair separations from the NVSS catalogue, \cite{2016ApJS..224...18P} produced a list of GRG candidates with $LAS \ge 240''$. From Figure~\ref{fig:GRGNOM_parameter} it can be seen that all of the GRG class objects used in this work are smaller than the limit of \cite{2016ApJS..224...18P}; however, it can also be seen that there is still a clear separation in $LAS$ between NOM class galaxies and GRGs and we suggest that it is for this reason that the inclusion of redshift information did not strongly affect the results of the models presented in this work. This $LAS$ separation is likely to be a consequence of the historic sample selection governing current catalogues of known GRGs. The development of labelled training sets with larger numbers of intermediate size radio galaxies will enable future studies to investigate this cross-over region of parameter space in more detail. 

\section{Conclusions}
\label{sec:conclusions}

Previous GRG searches have largely depended on visual inspection, and, while successful given the size of historic observational data bases, this human-powered methodology is unlikely to scale well to the new generation of radio sky survey data, which will require the investigation of millions of extended radio sources. 

In this work, we have explored the possibility of automated GRG identification through deep-learning by using 7 different CNN-based model architectures, including a multi-branched CNN algorithm that incorporates information from multiple surveys with different resolution. The best performing models in this study achieve \textcolor{black}{97.9\% and 91.4\%} test accuracy using the GRGNOM-A and GRGNOM-B test samples described in this work. \textcolor{black}{This result follows from an extensive investigation of model performance under different architectural choices, dataset shift and dataset composition.}

\textcolor{black}{A key result from this work is the introduction of multi-branched networks in order to boost model performance compared with the classical CNN architectures that use a single type of image input. By including host galaxy redshift information, model performance was also found to be improved. We find that importing both NVSS and FIRST images as dual inputs corrected 39\% of objects that were misclassfied by a single domain network.}

\textcolor{black}{Architecturally, IC as a regularization method was found to be non-harmful while need to be compared with other methods when deciding regularization strategies. Furthermore, it was shown that the use of the inception module could lower the model computational cost, affect model performance under certain circumstances, and was found to be able to correct misclassifications in the case when the target object is small and compact.}

Finally, we investigated the cause of frequent misclassifications by inspecting individual samples, and found that other than the aforementioned factors, a sample might be frequently misclassified if (a) its sample image contains multiple sources, (b) the standard pre-processing procedures have eliminated part of its extended morphology, or (c) its radio component was partly resolved out comparing with the survey map images used to identify the GRG (e.g., LoTSS at 151\,MHz). 

\section*{Acknowledgements}

The authors are grateful for the assistance of over 12,000 volunteers in the Radio Galaxy Zoo project, whose contributions are acknowledged at \url{http://rgzauthors.galaxyzoo.org}. The authors are also very grateful for discussions from the machine learning group at Jodrell Bank Centre for Astrophysics, JBCA. This research was supported by JBCA, University of Manchester. The corresponding author is also grateful for the effort of project team members, participating school teachers and students of RGZ\_CN, a teaching side project of the Radio Galaxy Zoo. AMS gratefully acknowledges support from an Alan Turing Institute AI Fellowship EP/V030302/1. This research made extensive use of Astropy\footnote{http://www.astropy.org}, a community-developed core Python package for Astronomy \citep{astropy:2013, astropy:2018}.

\section*{Data Availability}

\textcolor{black}{Exemplar model weights for all architectures presented in this work using IC as the regularization method are available from Zenodo (DOI: 10.5281/zenodo.5749316). The RGZ DR1 catalogue will be made publicly-available through Wong et al (2022; in preparation), at which time the GRGNOM-A/B datasets will also become available on Zenodo.}




\bibliographystyle{mnras}
\bibliography{ref.bib} 




\bsp	
\label{lastpage}
\end{document}